\documentclass[twocolumn]{aastex701}

\usepackage{amsmath}
\usepackage{amssymb}
\usepackage{amsfonts}
\usepackage{graphicx}
\usepackage{multirow}
\usepackage{bm}
\usepackage{hyperref}
\hypersetup{
  colorlinks=true,
  linkcolor=  blue,
  citecolor=  blue,
  urlcolor=   blue,
  anchorcolor=blue,
  linkcolor=  blue
  }
\usepackage[normalem]{ulem}
\usepackage{epsfig}
\usepackage{verbatim}
\usepackage{color}
\usepackage{mdframed}
\usepackage{soul} 
\usepackage{braket}
\usepackage{enumitem}
\usepackage[percent]{overpic}
\usepackage{float}
\usepackage{array,booktabs,tabularx}
\renewcommand{\arraystretch}{1.6}
\usepackage{soul}
\include{eps2pdf}
\usepackage{fontawesome}
\usepackage{sidecap}
\usepackage{multirow}
\usepackage[flushmargin,bottom]{footmisc}
\usepackage{longtable}
\usepackage[normalem]{ulem}
\usepackage{xspace}

\newcommand{\Ha}{\ensuremath{\text{H}\alpha}\xspace}
\newcommand{\oii}{\ensuremath{\text{[O\,II]}}\xspace}
\newcommand{\oiii}{\ensuremath{\text{[O\,III]}}\xspace}

\renewcommand{\thefootnote}{\fnsymbol{footnote}}

\shorttitle{Neural-network interloper mitigation for LIM}
\shortauthors{}

\begin{document}

\title{Disentangling Target Lines from Interlopers and Continuum with Neural Networks: \\ A SPHEREx Intensity Mapping Case Study}

\author{Marina S. Cagliari \footnotemark[1]} 
\affiliation{Laboratoire d’Annecy de Physique Theorique (LAPTh), CNRS/USMB, 99 Chemin de Bellevue BP110 - Annecy - \\ F-74941 - ANNECY CEDEX - FRANCE}\email{}

\author{Zucheng Gao \footnotemark[2]}
\affiliation{Laboratoire d’Annecy de Physique Theorique (LAPTh), CNRS/USMB, 99 Chemin de Bellevue BP110 - Annecy - \\ F-74941 - ANNECY CEDEX - FRANCE}\email{}

\author{Azadeh Moradinezhad Dizgah \footnotemark[3]}
\affiliation{Laboratoire d’Annecy de Physique Theorique (LAPTh), CNRS/USMB, 99 Chemin de Bellevue BP110 - Annecy - \\ F-74941 - ANNECY CEDEX - FRANCE}\email{}

\footnotetext[1]{\href{mailto:marina.cagliari@lapth.cnrs.fr}{\texttt{marina.cagliari@lapth.cnrs.fr}}}
\footnotetext[2]{\href{mailto:zucheng.gao@lapth.cnrs.fr}{\texttt{zucheng.gao@lapth.cnrs.fr}}}
\footnotetext[3]{\href{mailto:azadeh.moradinezhad@lapth.cnrs.fr}{\texttt{azadeh.moradinezhad@lapth.cnrs.fr}}}


\begin{abstract} 

Line-intensity mapping (LIM) traces the large-scale distribution of matter by measuring fluctuations in aggregate line emission from unresolved galaxies and the intergalactic medium, providing a powerful probe of both astrophysics and cosmology. However, interpreting LIM data is limited by our ability to disentangle the signal of a target spectral line from continuum emission and interloper lines, which are emissions from other redshifts that fall within the observed frequency band. Astrophysical modeling uncertainties further complicate matters, leaving the relative amplitudes of the map components poorly understood. In this paper, we present a neural-network (NN) approach to separate the three map components at the level of the angular power spectrum, explicitly accounting for uncertainties in their relative amplitudes. As test cases, we generate SPHEREx-like maps with variable interloper line luminosities across multiple frequency channels, with and without pixel-wise scatter and continuum contributions. We find that cross-channel correlations are essential for robust NN performance when scatter is present. The NN exhibits a hierarchy in residual errors: brighter components yield smaller residuals, and the dimmest the highest. Without continuum emission, the network recovers the target power spectrum to within $2.5\%$, while partially correcting the interloper spectra. With continuum included, the NN accurately reconstructs the power spectra of the continuum and target line, within $2\%$ and $6\%$, respectively, but fails to recover those of the interlopers. Reducing pixel-level scatter further improves performance, lowering residual errors to $1\%$ (continuum) and $3\%$ (target line).

\end{abstract}



\section{Introduction\label{sec:introduction}}

Line-intensity mapping (LIM), targeting molecular and atomic transitions from the radio to the ultraviolet, is a powerful and cost-effective spectroscopic technique for mapping the cosmic large-scale structure over large cosmological volumes. In contrast to galaxy redshift surveys that detect individual sources above a flux limit, LIM measures spatial fluctuations in the integrated emission of specific lines from unresolved galaxy populations and the intergalactic medium (IGM). The known rest-frame frequency of each transition provides the radial (redshift) coordinate, enabling tomographic, three-dimensional maps of structure. 

In addition to the well-studied $21\,\mathrm{cm}$ hyperfine transition of neutral hydrogen \citep{Pritchard:2011xb,Liu:2019awk}, LIM targeting other spectral lines has attracted growing interest in recent years as a probe of both astrophysics and cosmology \citep{Kovetz:2017agg,Bernal:2022jap,Kovetz:2019BAAS,Chang:2019BAAS,Silva:2019hsh,Karkare:2022bai}. A number of completed, ongoing, and upcoming pathfinder surveys target numerous transitions across the electromagnetic spectrum: in the mm and sub-mm, COPSS \citep{Keating:2016pka}, mmIME \citep{Keating:2020wlx}, COMAP \citep{Cleary:2021dsp}, FYST \citep{CCAT-prime:2022qkj}, and SPT-SLIM \citep{Karkare:2021ryi} target CO rotational ladder, while CONCERTO \citep{CONCERTO2020}, TIME \citep{TIME}, FYST, EXCLAIM \citep{Ade:2019ril}, and TIM \citep{TIM} target [CII] fine structure line; at optical and near-infrared wavelengths, HETDEX \citep{Gebhardt:2021vfo} focuses on Ly$\alpha$ line, while SPHEREx \citep{Dore2018} will probe \oii, \oiii, H$\alpha$, and H$\beta$ lines. Beyond this generation of (pathfinder) surveys, proposals for future space and ground-based surveys with enhanced sensitivity and sky coverage are also being considered \citep{Cooray:2016hro,COMAP:2021nrp,Delabrouille:2019thj,Silva:2019hsh,Karkare:2022bai}.

As an astrophysical probe, LIM associated with emission from star-forming galaxies constrains population-averaged properties of galaxies, including the numerous faint ones below traditional survey flux limits; thereby informing the cosmic history of star formation and galaxy assembly~\citep{Silva:2014ira,Li:2015gqa,Mashian:2015JCAP,Breysse:2015saa,Breysse:2016opl,Brown:2019ApJ,Yue:2015sua,Breysse:2019MNRAS} and the physics of reionization~\citep{Silva:2013ApJ,Pullen:2013dir,Comaschi:2015waa,Comaschi:2016soe,Sato-Polito:2020qpc,Zhou:2020arXiv,Zhou:2020hqh}. As a cosmological probe, LIM maps the LSS across a wide range of scales and redshifts, including ultra-large scales and high-redshift regimes where conventional galaxy surveys are challenging to perform. It therefore offer a powerful probe of fundamental physics by measuring expansion history of the Universe \citep{Karkare:2018sar,Bernal:2019gfq}, and the Baryon Acoustic Oscillations \citep[BAO;][]{Bernal:2019jdo,MoradinezhadDizgah:2021dei}, constraining primordial non-Gaussianity \citep{MoradinezhadDizgah:2018zrs,MoradinezhadDizgah:2018lac,Chen:2021ykb}, probing properties of cosmic neutrinos \citep{Bernal:2021ylz,MoradinezhadDizgah:2021upg,Shmueli:2024npx}, and shedding light on the nature of dark matter \citep{Creque-Sarbinowski:2018ebl,Bernal:2020lkd} and dark energy \citep[or possible modifications to gravity;][]{Scott:2022fev,MoradinezhadDizgah:2023src}.

Despite its considerable potential, LIM remains in an early observational phase, and interpreting current data is challenging due to uncertainties in theoretical modeling, the separation of line emission from continuum and line interlopers, and instrumental systematics from imperfect calibration. In this paper, we address the problem of disentangling the target line signal from the noise by line interlopers and extragalactic continuum foreground\footnote{In the frequency bands of SPHEREx and for angular scales best probed by the survey, the extra-galactic continuum dominates over the galactic continuum \citep{CIBER:2025aoi}.} for non–$21\,\mathrm{cm}$ LIM, with emphasis on optical and near-infrared lines accessible to the SPHEREx survey~\citep{Dore2018}.

Contamination by \emph{line interlopers} arises because, at many of the frequencies probed by non–$21\,\mathrm{cm}$ LIM surveys, multiple emission lines from different redshifts fall within a single spectral channel and contribute to the observed intensity. As a result, the measured fluctuations mix structure from distinct cosmic epochs. If left untreated, interlopers can degrade signal-to-noise and bias astrophysical and cosmological inferences \citep[e.g.,][]{Chen:2021ykb,Karkare:2022bai}. Developing efficient and accurate \emph{line deconfusion} methods is therefore essential for realizing the potential of non-radio LIM. Previously proposed strategies include: \emph{voxel} and \emph{source masking}, often informed by external galaxy catalogs, to remove the brightest, most contaminated voxels \citep{Silva2015,Sun2018,Bethermin:2022lmd,VanCuyck:2023uli}; \emph{cross-correlations with external tracers} at the target redshift, which select the desired signal while down-weighting uncorrelated interlopers \citep{Silva2013,Pullen2013,Yang:2019eoj}; \emph{multi-line} and \emph{multi-band cross-correlations} between intensity maps that isolate emission from the same redshift \citep{Visbal2010,Gong2012,Cheng:2020asz,Cheng:2022ani,Roy:2023pei,Cheng:2024nfy}; and \emph{power-spectrum anisotropy} methods that exploit the geometric distortion imprinted when interloper emission is mapped into the target-line comoving frame \citep{Cheng2016,Lidz2016}.

In addition to target line signal and interlopers, the observed voxel intensity in LIM surveys also receives contributions from \emph{continuum emission}, a bright, spectrally smooth component with both Galactic and extragalactic contributions, depending on the wavelength of interest, which can exceed the target signal by $1$ to $5$ orders of magnitude. Mitigation of bright Galactic synchrotron has been extensively studied for $21\,\mathrm{cm}$ using, for example, low-$k_\parallel$ masking and principal component analysis~\citep[PCA;][]{Matteo2002, Oh2003, Santos2005, Zaldarriaga2004, McQuinn2006, Liu2009, Chang2010, Liu2011, Chapman2012}. The extra-galactic continuum comes from a combination of stellar continuum, dust emission, and active galactic nuclei ~\citep[AGN;][]{Liu:2013}. Fewer studies address the removal of extragalactic continuum (e.g., \citealt{Gao-prep} in the near-IR and optical; \citealt{Silva2015} for far-IR $\mathrm{[C\,II]}$), but these indicate that PCA-based approaches can recover the underlying structure. 

Disentangling LIM signals from interlopers and continuum is further complicated by astrophysical uncertainties in modeling the components of the maps. Our understanding of processes relevant to galaxy formation and the IGM remains incomplete; different star-formation rate and metallicity prescriptions can produce order-of-magnitude variations in the predicted amplitudes of summary statistics of line fluctuations~\citep{Schaan:2021gzb}. In addition, scatter in the mass-to-luminosity relation can introduce extra small-scale power, and the effect is degenerate with cosmological parameters affecting small-scale clustering of dark matter distribution~\citep{Schaan:2021hhy,Sun2022}.  Further, the uncertainty on modeling dust, AGN, and stellar spectrum can influence the amplitude and shape of the extra-galactic continuum contamination~\citep{Conroy_2010}. 

In this paper, we adapt the machine-learning framework of \citet{Cagliari:2025xcq}, developed in the context of galaxy-clustering analysis, to the problem of disentangling the LIM signal from contributions of line interlopers and extragalactic continuum, explicitly accounting for uncertainties in the amplitudes of the various components. In slitless spectroscopic galaxy surveys with limited wavelength coverage that rely on a single emission line for redshift determination, interlopers are individual galaxies whose emission from a different rest-frame transition at another redshift is misidentified as the target line; such misclassifications distort the sample’s inferred redshift distribution and modify the clustering measurements, leading to biased cosmological constraints if unaccounted for. For example, in Euclid, both \oiii\ and $\mathrm{[S\,III]}$ emitters can be mistaken for \Ha, contaminating the target sample~\citep{Euclid:2025duk,Lee-prep}. It was shown in \citet{Cagliari:2025xcq} that a neural network (NN) based approach is effective in removing the line interloper contamination from the three-dimensional power spectrum monopole and in recovering the fraction of target and line interloper galaxies. 

Here, we transfer that approach to LIM and assess it using mock SPHEREx deep-field maps \citep{Gao-prep} that include extragalactic continuum emission and interloper lines. We consider three transitions, \Ha, \oiii, and \oii, and test robustness to astrophysical uncertainty via an overall mean-intensity scaling applied to the \oiii and \oii maps and a pixel-level log-normal scatter applied to all maps. We compare different input configurations that either use only auto-correlation information or combine auto- and cross-correlations. Additionally, we evaluate the method's performance in the presence of a bright continuum. We limit ourselves to component separation at the level of the angular two-point functions.

The rest of the paper is organized as follows. In section~\ref{sec:simulations}, we give a brief description of the construction of the simulated intensity maps. In section~\ref{sec:data_setup}, we describe the parameter specification and the three data configurations. In section~\ref{sec:NN}, the design of the NN and evaluation are discussed. Then we present our results in section~\ref{sec:results}, while we draw our conclusions in section~\ref{sec:conclusions}. The code used in this work is public.\footnote{\url{https://github.com/mcagliari/NoInterNet/tree/LIM}.}

\section{Simulation Setup}
\label{sec:simulations}

We analyze simulated intensity cubes constructed for the SPHEREx deep fields \citep{Gao-prep}. The pipeline begins with halo catalogs from $N$-body simulations; each halo is assigned a spectral energy distribution from which line luminosities are derived and projected into observed-frequency cubes. The simulations adopt the deep-field footprint and redshift coverage, include extragalactic continuum emission, and the full set of relevant interloper lines. Here, we do not apply an instrument model (beam, line-spread function, spectral sampling, and noise). Below, we summarize the main ingredients and assumptions; further details are provided in \citet{Gao-prep}.

\subsection{Dark Matter Halo Lightcone}
\label{sec:DM_lightcone}

Our simulations are based on the Hidden Valley $N$-body simulation suite~\citep{Modi2019}, which provides the underlying dark matter distribution. The suite includes 12 snapshots spanning redshifts from $z = 0.5$ to $z = 6.0$ in steps of $\Delta z = 0.5$. It assumes a $\Lambda$CDM model with $\Omega_{\rm M} = 0.309167$, $\Omega_{\rm b}=0.04903$, $h=0.677$, $n_s=0.96824$, and  $\sigma_8=0.8222$. Each simulation box has a comoving size of $1024\, \mathrm{Mpc}/h$, a particle mass resolution of $8.58\times 10^{7}\, \mathrm{M_\odot}/h$, and a comoving spatial resolution of $0.1\, \mathrm{Mpc}/h$. 

\begin{figure*}[t]
    \centering
    \includegraphics[width=\linewidth]{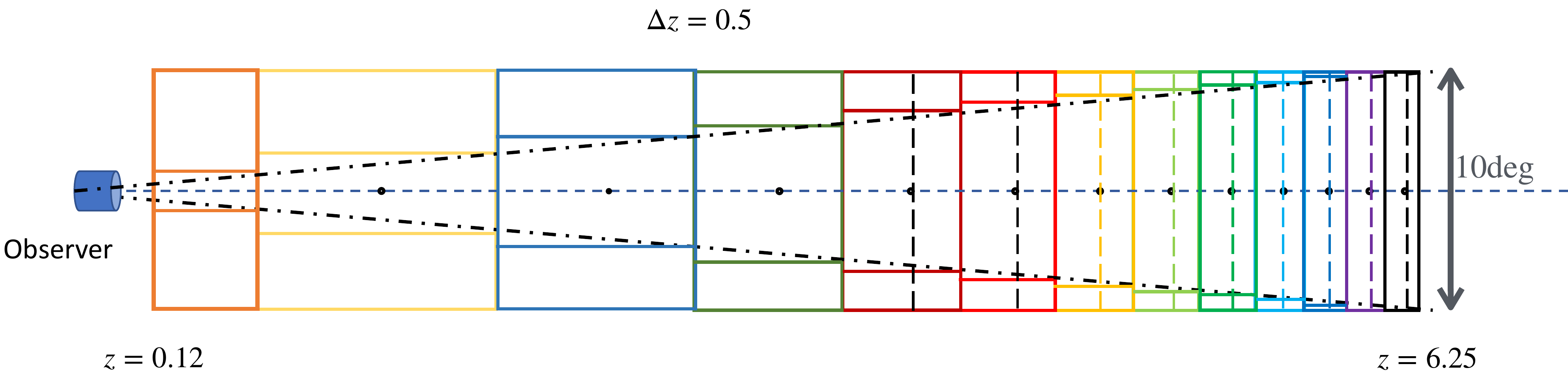}
    \caption{The design of the lightcone. We only select dark matter halos within the black dash-dotted line.}
    \label{fig:lightcone_plan}
\end{figure*}

We cut, rotate, and stitch the simulation boxes to construct a dark matter halo lightcone, as visualized in figure~\ref{fig:lightcone_plan}. The redshift range of the lightcone is $z=0.14\sim 6.25$. This results in an angular range of $10\times 10 \,\mathrm{deg}^2$, which matches the deep-field survey of the SPHEREx mission~\citep{Dore2018}. 

\begin{figure*}[htbp!]
    \centering
    \includegraphics[width=0.9\linewidth]{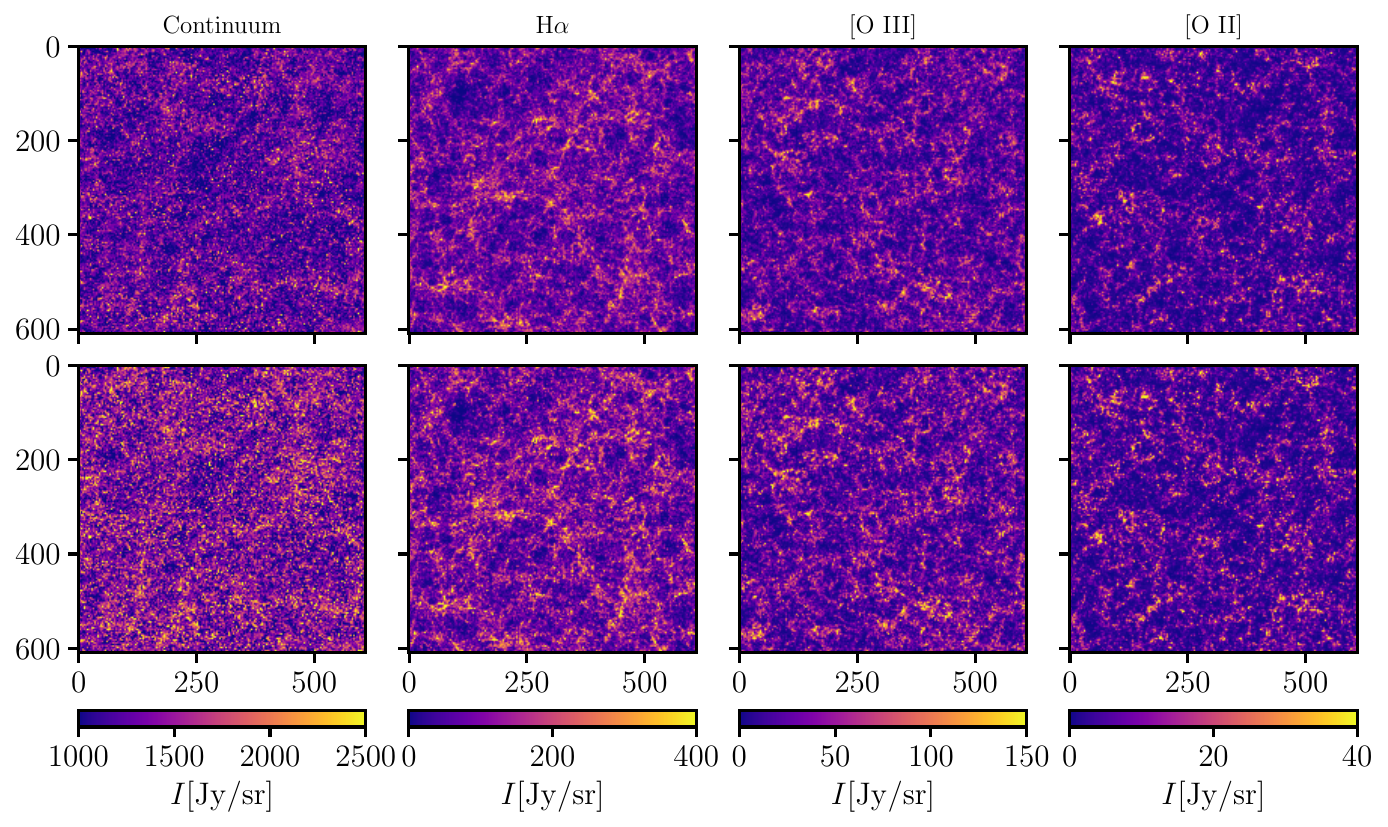}
    \caption{Simulated intensity maps in SPHEREx channel $27$, for continuum-only (first column), \Ha (second column), \oiii (third column),  and \oii(fourth column). The top row shows the baseline maps; the bottom row applies pixel-wise log-normal scatter with $\sigma=0.2 \, \rm dex$. See section~\ref{subsec:astrophysical_uncertainty} for details.}\vspace{0.15in}
    \label{fig:intensity_maps}
\end{figure*}

\subsection{Intensity Maps}
\label{sec:intensity_maps}

To include the interlopers and extra-galactic continuum, we assign a full spectral energy distribution (SED) to each dark matter halo. The SEDs are generated through the following three steps:
\begin{enumerate}
    \item To each halo, given its mass and redshift, we assign a star formation history
    using UniverseMachine~\citep{Behroozi_2019}. 
    \item We pass the star-formation history to the flexible stellar population synthesis (FSPS) code~\citep{Conroy_2009, Conroy_2010}, with the default astrophysical parameters.
    \item We obtain the SED as a function of halo mass and redshift, noted as $L_{\nu}(M_h, z)$.
\end{enumerate}

Given the SEDs of various lines, for each frequency channel $\mathrm{ch} = [\nu_0, \nu_1]$, the intensity contribution of an individual source within a pixel (with solid angle ${\rm d}\Omega_{\rm o}$) is given by
\begin{equation}\label{eq:intensity}
\begin{aligned}
&I_{[\nu_0, \nu_1]}(M_h,z_{\rm cosmo}) =  \frac{1}{{\rm d}\Omega_{\mathrm{o}}} \frac{1}{4\pi \chi^2} \frac{(1+z_{\mathrm{cosmo}})^2}{(1+z_{\mathrm{tot}})^4} \\
&\hspace{1.5cm} \times \int_{\nu_0(1+z_{\mathrm{tot}})}^{\nu_1(1+z_{\mathrm{tot}})}{\rm d}\nu_{\mathrm{e}} L_\nu(\nu_{\mathrm{e}}) f\left(  \frac{\nu_{\mathrm{e}}}{(1+z_{\mathrm{tot}})} \right) \, ,
\end{aligned}
\end{equation}
where $1+z_{\mathrm{tot}} = (1+z_{\mathrm{cosmo}})(1+z_{\mathrm{pec}})$ taking both the cosmological and kinetic redshift into account (to include redshift-space distortion), $\chi$ is the radial comoving distance of the source and $f(\nu)$ is the band-pass filter of the instrument. In this simulation, we choose $1\times 1 \, \mathrm{arcmin}^2$ as the pixel angular resolution, which is combining $10\times 10$ SPHEREx pixel, respectively. \vspace{0.1in}

We generate three kinds of intensity maps:
\begin{itemize}
    \item \textbf{All Emission}: it is the sum over the intensities of all sources along the line-of-sight of the pixel.
    \item \textbf{Continuum only}: from FSPS, we generate the continuum-only SED (no line emission), then follow the same procedure as the ``All Emission" case to generate maps.
    \item \textbf{Line}: for a line with rest-frame frequency $\nu_{\mathrm{line}}$, we only sum the sources within redshift bin $[\nu_{\mathrm{line}}/\nu_1-1, \nu_{\mathrm{line}}/\nu_0-1]$.
\end{itemize}

For this work, we analyze the Continuum-only component together with three emission-line maps, \Ha, \oii, and \oiii.  Figure~\ref{fig:intensity_maps} displays the simulated maps for the 27th frequency channel ($\mathrm{ch}_{27}$): the top row shows the baseline maps, and the bottom row includes a pixel-wise lognormal scatter used to model astrophysical variability. We will describe how we effectively simulate the astrophysical uncertainty in section~\ref{subsec:astrophysical_uncertainty}.

From the maps, we notice that the \Ha is the brightest line, followed by \oiii and then \oii, and the continuum map is about one order of magnitude larger than the brighter line. Note that in this work, we consider only the extragalactic continuum, but the galactic continuum is also a bright component in the map.
However, for the SPHEREx deep field, the galactic diffuse light starts to dominate the angular statistics at $\ell<10^3$~\citep{CIBER:2025aoi}, while the major constraining power comes from angular scales smaller than $10~\mathrm{arcmin}$, where the extra-galactic source dominates.\footnote{We thank Yun-Ting Cheng for highlighting this point.} Therefore, for this work, we decided to neglect the effect of the Galactic continuum, which we will consider in future work. The observed map will be the sum of all maps in each row. The goal of this work is then to reconstruct the angular power spectra of each separate map given the total angular power spectrum.

For different channels, the continuum maps are the projection of the same structures along the line-of-sight, while the lines trace different structures from different redshifts. 

\subsection{Astrophysical Uncertainty}
\label{subsec:astrophysical_uncertainty}

We effectively study two kinds of astrophysical uncertainties: first, the uncertainty in the mean line intensity; second, the uncertainty in the scattering of the individual source. 

In the original simulation, we assign a fixed SED model depending on halo mass and redshift. However, different star-formation rate density models~\citep{Gong_2017} and metallicity models will cause orders of magnitude mean intensity differences for different lines~\citep{Schaan:2021hhy, Schaan:2021gzb}. In this work, we simulate the effect of metallicity uncertainty by fixing the \Ha luminosity model to the default, and randomly scale the \oiii and \oii maps by scaling factors, $f_\oii$ and $f_\oiii$.
 
The second uncertainty comes from the fact that the luminosity-mass relation has scatter. Instead of adding a scatter in the relation between halo mass and line luminosities at a fixed redshift, as done, e.g., in ~\citet{MoradinezhadDizgah:2021dei}, we include the log-normal scatter in the relation between the observed intensity and line luminosity model at the pixel level, i.e.,
\begin{equation}
\label{eq:scattered_map}
    I_{\rm scattered} (\theta, \phi) = I (\theta, \phi) \exp(\epsilon_{\sigma}(\theta, \phi)),
\end{equation}
where $(\theta, \phi)$ is the celestial coordinate, and $\epsilon_{\sigma}(\theta, \phi)\sim \mathcal{N}(0, \sigma^2)$ is a random field drawn from a normal distribution. 

In summary, the total intensity map, including both types of astrophysical uncertainties, is modeled as
\begin{align}
    \label{eq:full_map_uncertain}
    I^{\rm ch}_{\rm full} (\theta, \phi) &= I^{\rm ch}_{\rm Cont, scattered} (\theta, \phi) + I^{\rm ch}_{\rm H\alpha, scattered} (\theta, \phi) \nonumber \\
    &+ f_{\rm[O\, III]} I^{\rm ch}_{\rm [O\,III], scattered} (\theta, \phi) \nonumber \\
    &+ f_{\rm[O\, II]}I^{\rm ch}_{\rm [O\,II], scattered} (\theta, \phi) \, ,
\end{align}
where the log-normal scattering level is assumed to be the same for all maps.

\subsection{Summary Statistics} 
\label{sec:summary-stats}

The target observable of this work is the angular power spectra. Provided the angular span of the map is small ($10\, \mathrm{deg}$), we apply the flat-sky approximation to estimate the angular power spectrum from maps. We define
\begin{equation}
     \langle I^{\rm ch_i}(\mathbf{\ell}) I^{\rm ch_j}(\mathbf{\ell}') \rangle= (2\pi)^2 \delta^{\rm D}\left(\mathbf{\ell}+\mathbf{\ell}'\right)C^{ij}_{\ell},
\end{equation}
where the $I^{\rm ch_i}(\mathbf{\ell})$ is the two-dimensional Fourier transformation of the two-dimensional map, $I^{\rm ch}(\theta, \phi)$, where $\mathbf{\ell}= 2\pi/\sqrt{\theta^2 + \phi^2}$ under flat-sky approximation. Here, $i$ and $j$ index frequency channels: $i=j$ denotes an auto–power spectrum, while $i\neq j$ denotes a cross–power spectrum (i.e., $P_{ii}$ is auto and $P_{ij}$ with $i\neq j$ is cross).

We bin the angular multipoles with width $\Delta\ell=50$ into $100$ bins. But as will be described later, in practice, we drop the first bin since it is dominated by cosmic variance. 

\section{Data setup}
\label{sec:data_setup}

In this section, we detail the datasets used to train the neural network, starting from the inclusion or omission of continuum emission, to the scaling factors applied to interloper maps, the choice of scatter model, and the selection of auto- and cross-channel spectra. The full set of configurations is summarized in Table~\ref{tab:res-OII}.

\subsection{Choices of astrophysical parameters}
\label{sec:astro-params}

We use two independent set of scaling factors $f_{\rm[O\, III]}$ and $f_{\rm[O\, II]}$. To generate them, we draw $1024$ samples from a two-dimensional Sobol sequence in the range $[0.1, 2]$. This choice avoids the intensity of \oiii and \oii being brighter than the target \Ha.

We model the pixel-wise scattering by applying a log-normal scatter to all map components with three types of standard deviation $\sigma$, $0\, \rm dex$ (no scattering), $0.1\, \rm dex$, and $0.2\, \rm dex$.

\subsection{Choices of spectral channels}
\label{sec:ch-choice}

We consider three configurations for the input data vector: first, single-channel (SC), using only the auto–power spectrum of a target frequency channel, ${\rm ch}^{\rm tar}$; second, multi-channel (MC), augmenting SC with companion channels ${\rm ch}^{\rm inter}_i$ and the corresponding cross–power spectra; and, third, multi-channel plus continuum (MC+C), further including a cross–power with a continuum-dominated channel.

For a given spectral channel, the redshifts at which various lines contribute are known. For the target channel ${\rm ch}^{\rm tar}$, we denote the redshift at which the $i$-th line contributes as $z_i^{\rm tar}$. For each interloper $i$ we select a matched channel ${\rm ch}^{\rm inter}_i$ in which the dominant line appears at the same redshift as the $i$-th interloper in the target channel.  The MC data vector therefore includes the auto–power spectra of ${\rm ch}^{\rm tar}$ and all ${\rm ch}^{\rm inter}_i$, together with the cross–power spectra $\mathrm{ch^{tar}} \times \mathrm{ch}^{\rm inter}_{i}$. These cross-spectra are dominated by structure at $z_i^{\rm tar}$ and are commonly used for interloper mitigation in LIM~\citep{Cheng:2024nfy}.

\begin{figure}
    \centering
    \includegraphics[width=0.9\linewidth]{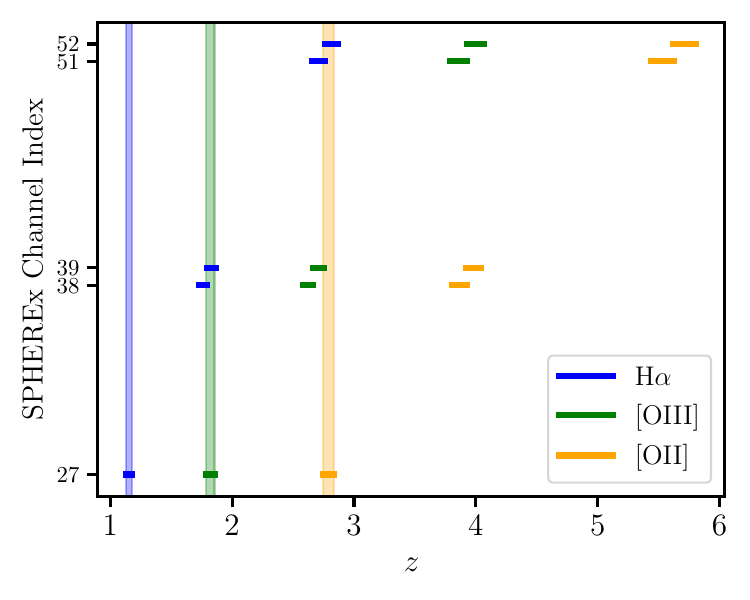}
    \caption{The redshifts of lines in target channels and matched channels. The blue, green, and orange bands are the redshift bins of \Ha, \oii, and \oiii in several spectral channels. }
    \label{fig:interloper_channel}
\end{figure}

In this work, we take \Ha as the target line and treat \oii and \oiii as interlopers. The target channel is set to $\rm ch_{27}$ of SPHEREx mission, corresponding to \Ha located at $z=1.13$. The matched interloper channels are $\rm ch_{39}$ for \oiii and $\rm ch_{52}$ for \oii (see figure~\ref{fig:interloper_channel}). While additional matches are possible (every channel has its own set of interlopers), we restrict attention to these primary pairs: at higher-order matches, the relevant interlopers shift to progressively higher redshift and fainter intensity, and the required channels approach or leave the instrument bandpass.

When continuum emission is included, we adopt the MC+C configuration by adding the cross–power $\rm ch^{tar} \times ch^{cont}$, where $\text{ch}^{{\rm cont}}$ is chosen such that no line in that channel lies at the same redshift as any line contributing to $\rm ch^{tar}$. This cross-spectrum is therefore dominated by the continuum common to both channels and provides additional leverage to disentangle the continuum contribution. In our tests $\rm ch_{34}$ serves as ${\rm ch}^{\rm cont}$.

For cases including continuum, we also test a residual-continuum scenario by rescaling the continuum \emph{power spectrum} by a factor of 0.02, representative of percent-level residuals after subtracting the first few PCA modes in map-level cleaning \citep{Gao-prep}.

\section{Neural Network Design and Evaluation}
\label{sec:NN}

In this section, we describe the neural network we utilize to extract different components of the angular power spectrum and how we pre-process the data.

\subsection{Network Description}
\label{sec:nointernet}

\begin{table*}[t]
\caption{\label{tab:architecture}Specifications of the neural network architecture. For each {\it Map} configuration, the first row in the {\it Input} column corresponds to SC case, and the second row to MC or MC+C cases described in section~\ref{sec:ch-choice}.}
\begin{ruledtabular}
\begin{tabular}{cccccc}
Map  & Input & Input size & $n_{\rm in}$ & $n_{\rm min}$ & $n_{\rm out}$ \\ \midrule
\multirow{3}{*}{$\Ha + \oii + \oiii$} 
     & $\text{ch}_{27} \times \text{ch}_{27}$ & 99 & 64 & 16 & 256 \\ \cmidrule(l){2-6}
     & $\{\text{ch}_{27} \times \text{ch}_{27}, \text{ch}_{52} \times \text{ch}_{52}, \text{ch}_{39} \times \text{ch}_{39},$ & \multirow{2}{*}{594} & \multirow{2}{*}{64} & \multirow{2}{*}{16} & \multirow{2}{*}{256}  \\
     & $\text{ch}_{27} \times \text{ch}_{52}, \text{ch}_{27} \times \text{ch}_{39}, \text{ch}_{39} \times \text{ch}_{49} \}$ \\ \midrule
\multirow{4}{*}{$\Ha + \oii + \oiii + {\rm Cont.}$} 
    & $\{\text{ch}_{27} \times \text{ch}_{27}, \text{ch}_{52} \times \text{ch}_{52}, \text{ch}_{39} \times \text{ch}_{39},$ & \multirow{2}{*}{594} & \multirow{2}{*}{64} & \multirow{2}{*}{16} & \multirow{2}{*}{256} \\
    & $\text{ch}_{27} \times \text{ch}_{52}, \text{ch}_{27} \times \text{ch}_{39}, \text{ch}_{39} \times \text{ch}_{49} \}$ \\ \cmidrule(l){2-6}
    & $\{\text{ch}_{27} \times \text{ch}_{27}, \text{ch}_{52} \times \text{ch}_{52}, \text{ch}_{39} \times \text{ch}_{39}, \text{ch}_{27} \times \text{ch}_{52},$ & \multirow{2}{*}{693} & \multirow{2}{*}{64} & \multirow{2}{*}{16} & \multirow{2}{*}{256} \\
    & $ \text{ch}_{27} \times \text{ch}_{39}, \text{ch}_{39} \times \text{ch}_{49}, \text{ch}_{27} \times \text{ch}_{34} \}$ \\
\end{tabular}
\end{ruledtabular}
\vspace{0.16in}
\end{table*}

We build on the fully connected architecture of \citet{Cagliari:2025xcq}, originally developed for spectroscopic galaxy surveys, and modify it to meet the requirements of line–intensity mapping. Below, we summarize the model and highlight the changes relative to the spectroscopic galaxy survey case that are specific to LIM.

We employ a moment network architecture \citep{Jeffrey:2020itg,CAMELS:2021raw} that takes as input the contaminated statistics (see section~\ref{sec:ch-choice} for details on the input features) and outputs the angular power spectrum of different components (\Ha, \oii, \oiii), the continuum when present, and the rescaling applied to the interloper luminosity. The moment network offers the advantage of providing corresponding estimated errors in addition to the mean values of the above quantities. In summary, in the case without the continuum, we define the output vector as
\allowdisplaybreaks{
\begin{align}
    \mathbf{y} = \{ & \mathbf{y}_{\rm c}^{\Ha}, \mathbf{y}_{\rm c}^{\oii}, \mathbf{y}_{\rm c}^{\oiii}, y_{\oii}, y_{\oiii}, \notag \\
    & \bm{\sigma}_{\rm c}^{{\rm H}\alpha}, \bm{\sigma}_{\rm c}^{\oii}, \bm{\sigma}_{\rm c}^{\oiii}, \sigma_{\oii}, \sigma_{\oiii}\},  
    \label{eq:output} 
\end{align}}
where the bold face symbols represent vector-like outputs. Here, $\mathbf{y}_c^{\rm comp}$ denotes the first moment (mean) of the correction vector to be applied to the contaminated angular power spectrum in order to recover the contribution of the component indicated by the superscript. We define their true values as follows
\begin{equation}
    \widehat{\mathbf{y}}_{\rm c}^{\rm comp} = \frac{C^{\rm comp}_\ell}{C^{\rm contam}_\ell} \, ,
    \label{eq:ycorr} 
\end{equation}
with $C^{\rm comp}_\ell$ the angular power spectrum of one of the components and $C^{\rm contam}_\ell$ the angular power spectrum of the map with all the components (All Emission) in the frequency channel of interest ($\text{ch}^{\text{tar}}$). Then, $y_{\oii}$ and $y_{\oiii}$ are the predicted scaling factors applied to \oii and \oiii, respectively. Their true values are $f_{\oii}$ and $f_{\oiii}$. Finally, in the second row of equation~\eqref{eq:output}, $\bm{\sigma}_{\rm c}^{\rm comp}$, $\sigma_{\oii}$, and $\sigma_{\oiii}$ are the second moments of the correction to get the angular power spectrum of one of the components and the interloper scaling factors. In the case with the continuum, the definition of equation~\eqref{eq:output} will have two additional terms $\mathbf{y}_{\rm c}^{{\rm cont}}$ (the true value of which is defined following equation~\ref{eq:ycorr}) and $\bm{\sigma}_{\rm c}^{{\rm cont}}$, which represents the first and second moments of the correction to get the continuum angular power spectrum, respectively. The total dimension of the output for the cases without and with the continuum is $598$ and $796$, respectively. As for the input of the network, we describe it in section~\ref{sec:ch-choice} and summarize them in the second column of Table~\ref{tab:architecture}.

As in \citet{Cagliari:2025xcq}, the network is a stack of dense layers that first applies a compression to the data and then decompresses it. Except for the last layer, the others use the LeakyReLu as activation function \citep{Maas2013RectifierNI}. The compression block of the network reduces the input dimension to $n_{\rm in}$ neurons and then progressively halves the layer size down to $n_{\rm min}$. The decompression progressively doubles the layers' dimension from $n_{\rm min}$ up to $n_{\rm out}$ and has a last layer that produces an output of the correct final size, which is the dimension of the output vector $\mathbf{y}$ (see above). In Table~\ref{tab:architecture}, we summarize the key information about the architecture of the network in the different configurations, and explicitly write the input features we provide to the network depending on the map components. 

We modify the loss function of \citet{Cagliari:2025xcq}, the standard objective used for moment networks \citep{Jeffrey:2020itg,CAMELS:2021raw}, by adding a constraint that forces, in each $\ell$-bin, the component corrections sum to unity. This enforces that the sum of the recovered component power spectra equals the power spectrum of the combined map. Strictly, one should also include cross–correlation terms between components; however, because the lines arise from widely separated redshifts, these terms are negligible compared to the auto-correlations. Incorporating this prior into the objective yields a physically informed neural network \citep[PINN;][]{2019JCoPh.378..686R}. Therefore, our one-batch loss is
\begin{widetext}
\allowdisplaybreaks{
\begin{align}
        L(\mathbf{y}, \mathbf{\widehat{y}}) & = \log \left( \frac{1}{n_b} \sum_{i = 1}^{n_b} \left( \sum_{j = 1}^{n_\ell} \sum_{\rm comp} \left( \widehat{y}_{{\rm c}, ij}^{\rm comp} - y_{{\rm c}, ij}^{\rm comp} \right)^2 \right) \right) + \log \left( \frac{1}{n_b} \sum_{i = 1}^{n_b} \sum_{\rm int} \left( f_{{\rm int}, i} - y_{{\rm int}, i} \right)^2 \right) \notag \\
        & + \log \left( \frac{1}{n_b} \sum_{i = 1}^{n_b} \left( \sum_{j = 1}^{n_\ell} \sum_{\rm comp} \left ( \left( \widehat{y}_{{\rm c}, ij}^{\rm comp} - y_{{\rm c}, ij}^{\rm comp} \right)^2 - \left( \sigma_{{\rm c}, ij}^{\rm comp} \right)^2 \right)^2 \right) \right) + \log \left( \frac{1}{n_b} \sum_{i = 1}^{n_b} \sum_{\rm int} \left( \left( f_{{\rm int}, i} - y_{{\rm int}, i} \right)^2 - \sigma_{{\rm int}, i}^2 \right)^2 \right) \notag \\
        & + \lambda \log \left( \frac{1}{n_b} \sum_{i = 1}^{n_b} \left( \sum_{j = 1}^{n_\ell} \left( \left( \sum_{\rm comp} Y_{{\rm c}, ij}^{\rm comp} \right) - 1 \right)^2 \right) \right) \, ,
    \label{eq:loss}
\end{align}}
\end{widetext}
where $\widehat{\mathbf{y}} = \{ \widehat{\mathbf{y}}_{\rm c}^{{\rm H}\alpha}, \widehat{\mathbf{y}}_{\rm c}^{\oii}, \widehat{\mathbf{y}}_{\rm c}^{\oiii}, f_{\oii}, f_{\oiii} \}$, with $\widehat{\mathbf{y}}_{\rm c}^{\rm cont}$ when the continuum is present, $n_b$ is the batch size (we set $n_b=64$ throughout), and $n_\ell$ is the number of $\ell$-bins of the angular power spectrum. The loss function additionally involves summations over the map components, ${\rm comp} = \{ {\rm H}\alpha, \oii, \oiii \}$, plus continuum when present, and the interloper lines, ${\rm int} = \{ \oii, \oiii \}$. The last term imposes the physical constraint described above; $\lambda$ sets its relative weight with respect to the other elements of the loss function (we use $\lambda=0.05$). We define $Y_{{\rm c}, ij}^{\rm comp} = \Phi^{-1} (y_{{\rm c}, ij}^{\rm comp})$, where $\Phi^{-1}$ is the inverse of the pre-processing normalization applied to the labels (see  section~\ref{sec:proc-metrics}). This transformation is applied because the unity-sum constraint is formulated for the \emph{unnormalized} power spectra. For brevity, we reuse the same symbols for normalized quantities (i.e., we abuse notation and write $\widehat{\mathbf{y}}$ after applying $\Phi$).

We use the adaptive moment estimation optimizer \citep[Adam;][]{kingma2014adam} with starting learning rate $l_r = 0.001$ to train the network. For the training, we set a maximum number of epochs of $2000$ and we impose an early stopping mechanism based on the validation loss function with a patience of $500$.

\subsection{Data Processing and Performance Metrics}
\label{sec:proc-metrics}

In our analysis, we test the algorithm on five different datasets: maps with only line contribution without any scatter, maps with only line contribution and scatter in the maps, and maps with line and continuum contributions with two different scatter levels and in one scatter case with lower continuum contribution (see section~\ref{sec:simulations}). All these datasets contain $1024$ maps, and we split them into training, validation, and test sets, which comprise $75$, $15$, and $10\%$ of the data. 

We normalize all the input features and the labels of the network in the range $[0,1]$. For a vector $\mathbf{a}$ this normalization is performed element-wise,
\begin{equation}
    a_i^{\rm n} = \frac{a_i - \min_j \left( A_{ij} \right)}{\max_j \left( A_{ij} \right) - \min_j \left( A_{ij} \right)} \equiv \Phi (a_i)\, ,
    \label{eq:norm}
\end{equation}
where $a_i$ is the $i$-th element of $\mathbf{a}$, $\mathbf{A}$ is the dataset of which $\mathbf{a}$ is part, and $j$ runs over the samples of the dataset.

We quantify the performance of the network using the same metrics of \citet{Cagliari:2025xcq}: the mean-squared error (${\rm MSE}$) and the reduced chi-squared ($\chi_{\rm red}^2$) for the first and second moments of the interloper scaling factor, and the mean correction error (${\rm MCE}$) for the first moment of the correction to the contaminated angular power spectrum to recover each component. We report the ${\rm MSE}$ and the $\chi_{\rm red}^2$ separately for the two interlopers (${\rm int} = \{ \oii, \oiii \}$),
\begin{align}
    {\rm MSE}_{\rm int} & = \frac{1}{n_{\rm test}} \sum_{i=1}^{n_{\rm test}} \left( f_{{\rm int}, i} - y_{{\rm int}, i} \right)^2 \, ,
    \label{eq:fMSE} \\ & \notag \\
    \chi_{\rm red, int}^2 & = \frac{1}{n_{\rm test}} \sum_{i=1}^{n_{\rm test}} \frac{\left( f_{{\rm int}, i} - y_{{\rm int}, i} \right)^2 }{\sigma_{{\rm int}, i}^2} \, .
    \label{eq:fchi2}
\end{align}
Similarly, we define the ${\rm MCE}$ for each components we aim to recover (${\rm comp} = \{ {\rm H}\alpha, \oii, \oiii, {\rm cont} \}$),
\begin{equation}
    {\rm MCE}_\ell^{\rm comp} = 1 - \frac{1}{n_s} \sum_{i=1}^{n_s} \frac{y_{{\rm c}, i}^{\rm comp}(\ell)}{\widehat{y}_{{\rm c}, i}^{\rm comp}(\ell)} \, ,
    \label{eq:MCE}
\end{equation}
where we are summing over the $n_s$ simulations that have the two scaling factors within some given ranges ($f_{\oii}^{\rm min} < f_{\oii} < f_{\oii}^{\rm max}$ and $f_{\oiii}^{\rm min} < f_{\oiii} < f_{\oiii}^{\rm max}$). Equation~\eqref{eq:MCE} gives an estimate of the residual error in the angular power spectrum of the recovered component. We do not define a specific statistic to quantify the error in the error bars outputted by the network; however, we perform error propagation to estimate the error in the ${\rm MCE}_\ell^{\rm comp}$, to visually evaluate if there is a residual bias in the correction. In general, we aim at a residual error below $2.5 \%$ for the \Ha and continuum contributions, and below $5 \%$ for the two interloper lines.

\section{Results}
\label{sec:results}

\begin{table*}[t]
\caption{\label{tab:res-OII}Scaling factor MSE and $\chi_{\rm red}^2$ for \oii for different configurations. The $\chi_{\rm red}^2$ values marked by * are computed removing the points with predicted uncertainty of $<10^{-3}$.}
    \begin{ruledtabular}
    \begin{tabular}{cccccccc}
        Map & Scatter [dex] & \multicolumn{2}{c}{Single-channel} & \multicolumn{2}{c}{Multi-channel} & \multicolumn{2}{c}{Multi-channel + Cont.} \\ \midrule
         &  & MSE & $\chi_{\rm red}^2$ & MSE & $\chi_{\rm red}^2$ & MSE & $\chi_{\rm red}^2$ \\ \midrule
         \multirow{2}{*}{$\Ha + \oii + \oiii$} & 0 & $1.45 \times 10^{-4}$ & $5.42^*$ & $3.26 \times 10^{-5}$ & $6.49^*$ & -- & -- \\ \cmidrule{2-8}
         & 0.2 & $1.74 \times 10^{-1}$ & $1.51$ & $1.23 \times 10^{-3}$ & $29.53^*$ & -- & -- \\ \midrule
         \multirow{2}{*}{$\Ha + \oii + \oiii + {\rm Cont.}$} & 0.2 & -- & -- & $3.02 \times 10^{-1}$ & $1.18$ & -- & -- \\ \cmidrule{2-8}
         & 0.1 & -- & -- & $6.50 \times 10^{-2}$ & $1.59$ & $5.57 \times 10^{-2}$ & $1.51$ \\ \midrule
         $\Ha + \oii + \oiii + {\rm resc. \, Cont.}$ & 0.1 & -- & -- & $4.75 \times 10^{-3}$ & $7.06$ & -- & -- \\
    \end{tabular} 
    \end{ruledtabular}
\end{table*}

\begin{table*}[t]
\caption{\label{tab:res-OIII}Same as Table~\ref{tab:res-OII}, but for \oiii.}
    \begin{ruledtabular}
    \begin{tabular}{cccccccc}
        Map & Scatter [dex] & \multicolumn{2}{c}{Single-channel} & \multicolumn{2}{c}{Multi-channel} & \multicolumn{2}{c}{Multi-channel + Cont.} \\ \midrule
         &  & MSE & $\chi_{\rm red}^2$ & MSE & $\chi_{\rm red}^2$ & MSE & $\chi_{\rm red}^2$ \\ \midrule
         \multirow{2}{*}{$\Ha + \oii + \oiii$} & 0 & $9.76 \times 10^{-5}$ & $3,56^*$ & $1.96 \times 10^{-5}$ & $0.57^*$ & -- & -- \\ \cmidrule{2-8}
         & 0.2 & $4.84 \times 10^{-3}$ & $1.18$ & $1.92 \times 10^{-4}$ & $3.24$ & -- & -- \\ \midrule
         \multirow{2}{*}{$\Ha + \oii + \oiii + {\rm Cont.}$} & 0.2 & -- & -- & $3.20 \times 10^{-2}$ & $1.80$ & -- & -- \\ \cmidrule{2-8}
         & 0.1 & -- & -- & $7.62 \times 10^{-3}$ & $2.24$ & $3.29 \times 10^{-3}$ & $6.57^*$ \\ \midrule
         $\Ha + \oii + \oiii + {\rm resc. \, Cont.}$ & 0.1 & -- & -- & $4.43 \times 10^{-4}$ & $1.94$ & -- & -- \\
    \end{tabular} 
    \end{ruledtabular}
\end{table*}

We present results for all configurations, without and with continuum and without and with pixel-wise scatter. Metrics and figures are evaluated on the test set of 102 realizations.

We compare configurations directly. A fully rigorous comparison would average over multiple training initializations for each case; here, we report single-training results for most configurations. Nevertheless, we believe that our results are statistically significant. In one representative setup (lines with reduced continuum contribution), we trained multiple networks and found stable MSE and MCE values across initializations. The reduced chi-squared, $\chi_{\rm red}^2$, exhibited more variability, so we interpret this metric more conservatively. 
 
\subsection{Lines-only}
\label{sec:lines}

First, we discuss the results from the simplified case with no continuum contribution. In this configuration, the maps only have three line components: the target line \Ha and the two interloper lines \oii and \oiii. In this configuration, the network is trained to recover the interloper scaling factors and the corrections to the target-channel auto–power spectrum needed to reconstruct the angular power spectrum of each component. As inputs, we test first the auto–power spectrum of the target channel alone (SC) and then the combination of auto– and cross–power spectra with the matched interloper channels (MC; see section~\ref{sec:ch-choice}).

\subsubsection{No scatter}
\label{sec:no-scatter-lines}

\begin{figure*}[htbp!]
    \centering
    \includegraphics[width=0.99\linewidth]{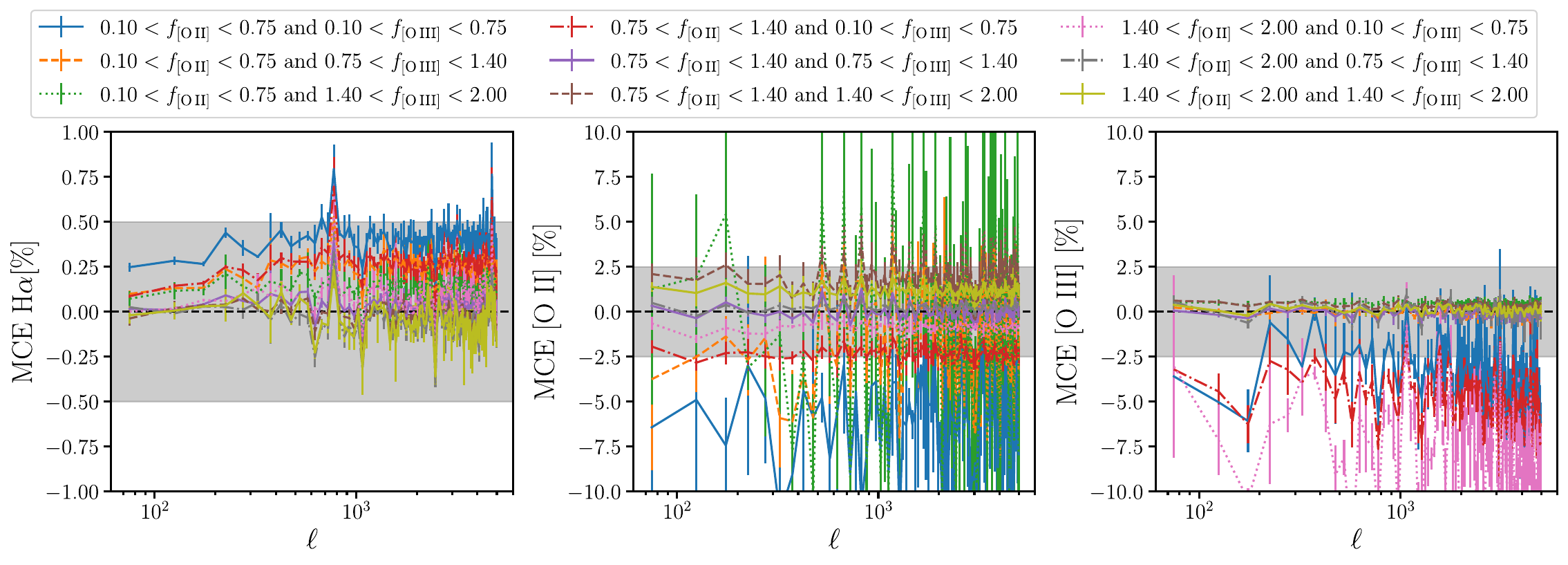}
    \caption{Mean correction error of angular spectra in the single-channel, line-only, no-scatter case. The shaded area indicates the $0.5\%$ error threshold for \Ha (left), the $2.5\%$ one for \oii (middle) and \oiii (right). The error bars correspond to the mean of the predicted uncertainty in each $\ell$-bin divided by the number of samples falling in the scale factor range.} \vspace{0.15in}
    \label{fig:line-NS-MCE-1ch}
\end{figure*}

\begin{figure*}[htbp!]
    \centering
    \includegraphics[width=0.99\linewidth]{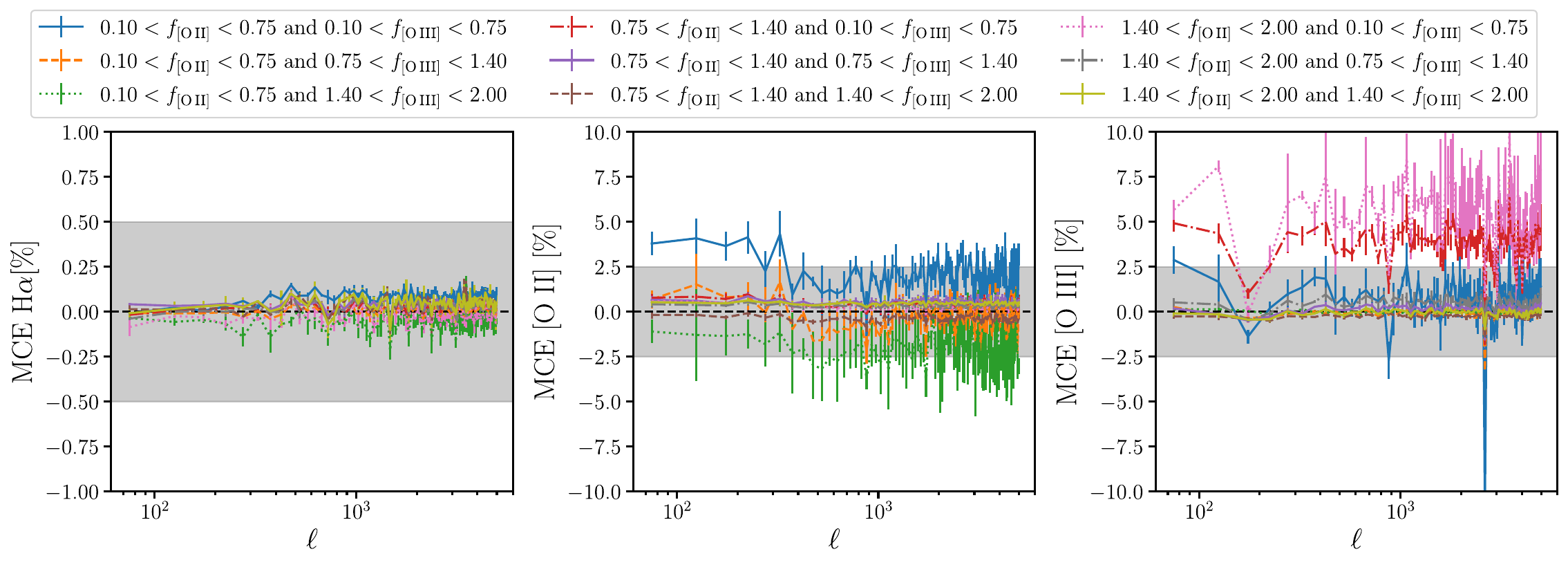}
    \caption{Same as figure~\ref{fig:line-NS-MCE-1ch}, but using multi-channel information.}
    \label{fig:line-NS-MCE-Mch}
\end{figure*}

\begin{figure*}[htbp!]
    \centering
    \includegraphics[width=0.99\linewidth]{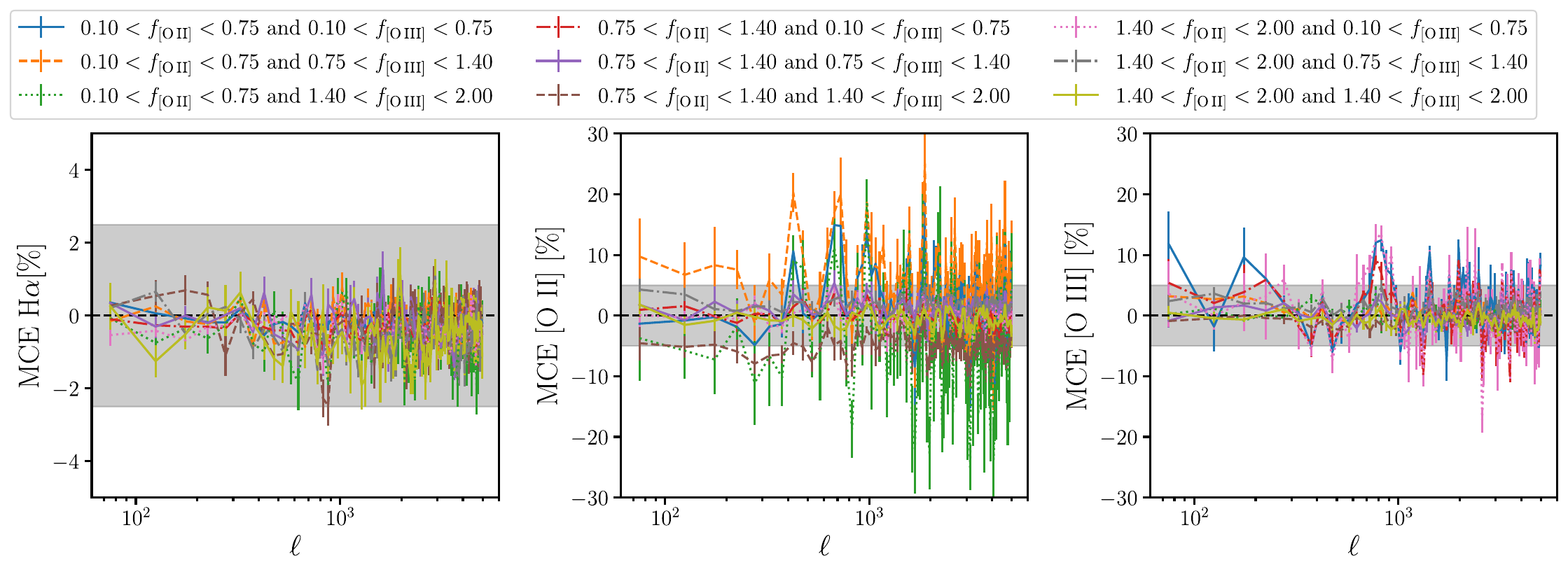}
    \caption{Same as figure~\ref{fig:line-NS-MCE-1ch} for the multi-channel analysis of line-only sample including a scatter of $0.2 \, {\rm dex}$. The shaded area corresponds to the $2.5\%$ error threshold for \Ha (left) and the $5\%$ one for \oii (middle) and \oiii (right).}\vspace{0.15in}
    \label{fig:line-S-MCE-Mch}
\end{figure*}

The most idealized setup contains only line emissions, with no added scatter. We use this case as a baseline to assess the network’s performance and its ability to recover the angular power spectra of the individual lines.

We first evaluate the performance of the network using the MSE and the $\chi_{\rm red}^2$ of the scaling factor, which we report in Tables~\ref{tab:res-OII} and \ref{tab:res-OIII} for \oii and \oiii, respectively. These metrics correlate with the fidelity of the recovered component angular power spectra, especially the MSE. Both input configurations, single-channel and multi-channel, yield low MSEs, with the multi-channel case typically improving performance by a factor of $\sim 4$--$5$. We also find that the MSE of the \oiii scaling factor is consistently lower than that of \oii; a similar hierarchy appears in the power-spectrum corrections. This trend is consistent with the network preferentially constraining the brighter component first (see discussion below).

At the $\chi_{\rm red}^2$ level, the network tends to be overconfident in its second moment estimation, except in the multi-channel \oiii case, where it overestimates the error bars. In Tables~\ref{tab:res-OII} and \ref{tab:res-OIII}, we marked with $*$ the $\chi_{\rm red}^2$ values that we computed removing the samples with an inferred second moment lower than $10^{-3}$ as we deem these predictions unreliable. We observe that in this case, all the input configurations produce extremely small second moments, and they can comprise up to $10\%$ of the test set. We do not speculate on the origin of this behavior here, noting only that it is confined to this configuration.\footnote{For all other analyses, there is at most one sample that does not satisfy the second moment threshold condition we imposed.} 

In figures~\ref{fig:line-NS-MCE-1ch} and \ref{fig:line-NS-MCE-Mch} we plot the mean correction errors for the single- and multi-channel inputs, respectively. Panels are ordered left-to-right as \Ha, \oii, and \oiii. Each curve corresponds to a joint bin in the \oii and \oiii scaling factors ($f_\oii,f_\oiii$).

In the single-channel case (figure~\ref{fig:line-NS-MCE-1ch}), across all interloper scaling–factor bins, the \Ha power spectrum is recovered to better than $1\%$ over all angular scales; performance is worst in the bins with the smallest interloper fractions (blue solid line), which is consistent with edge-of-prior effects on the scaling factors. For the interloper lines, in general, residuals are larger for the fainter component and decrease with line brightness. In the fiducial (unscaled) case, \oii is fainter than \oiii; in the bins shown in red dash-dotted and pink dotted lines, this hierarchy is reversed, and, as expected, the \oii spectrum is recovered more accurately than \oiii. Apart from these two bins, and the blue bin with the faintest interlopers, the \oiii power spectrum is recovered to better than $2.5\%$.

Including multiple channels (figure~\ref{fig:line-NS-MCE-Mch}) improves recovery of the \Ha power spectrum by more than a factor of $4$ (to better than $0.25\%$). Noticeably, including the additional channels results in recovering the \Ha power spectrum with accuracy largely independent of the scaling amplitude. For \oii, the residual error decreases by more than a factor of $2$ (to better than $5\%$). For \oiii, multi-channel information substantially improves performance in the lowest–fraction bin (blue solid curve), indicating reduced edge-of-prior effects. In bins where the \oii/\oiii brightness hierarchy is reversed (red dash-dotted and pink dotted lines), the additional channels do not markedly change the magnitude of the residuals but do flip their sign. 

\begin{figure*}[htbp!]
    \centering
    \includegraphics[width=0.9\linewidth]{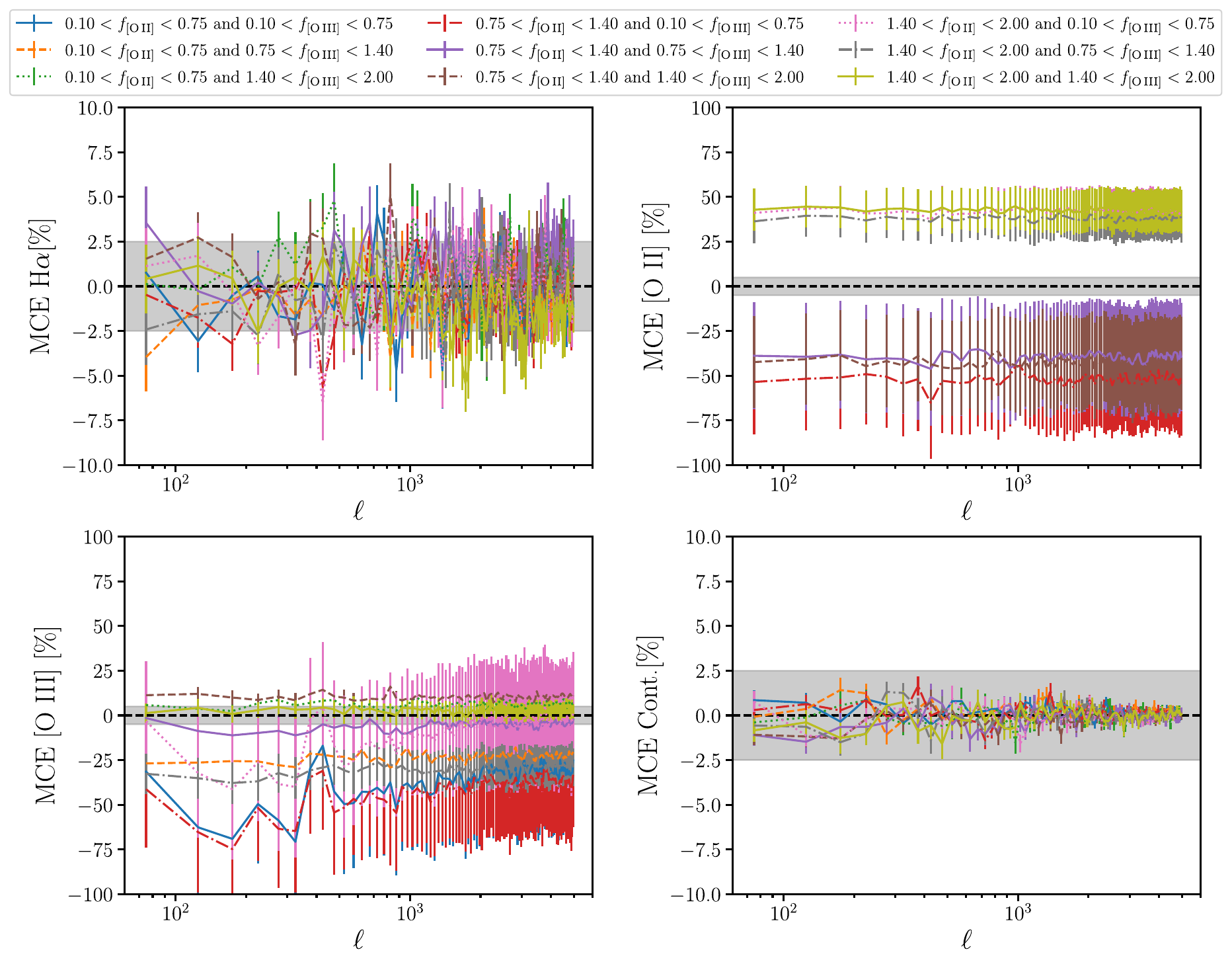}
    \caption{Same as figure~\ref{fig:line-NS-MCE-1ch} for the multi-channel analysis of the line-plus-continuum sample with scatter of $0.2 \, {\rm dex}$. The shaded area corresponds to the $2.5\%$ error threshold for \Ha (top left) and the continuum (bottom right), and the $5\%$ one for \oii (top right) and \oiii (bottom left).}\vspace{0.15in}
    \label{fig:all-S02-MCE-Mch}
\end{figure*}

\begin{figure*}
    \centering
    \includegraphics[width=0.9\linewidth]{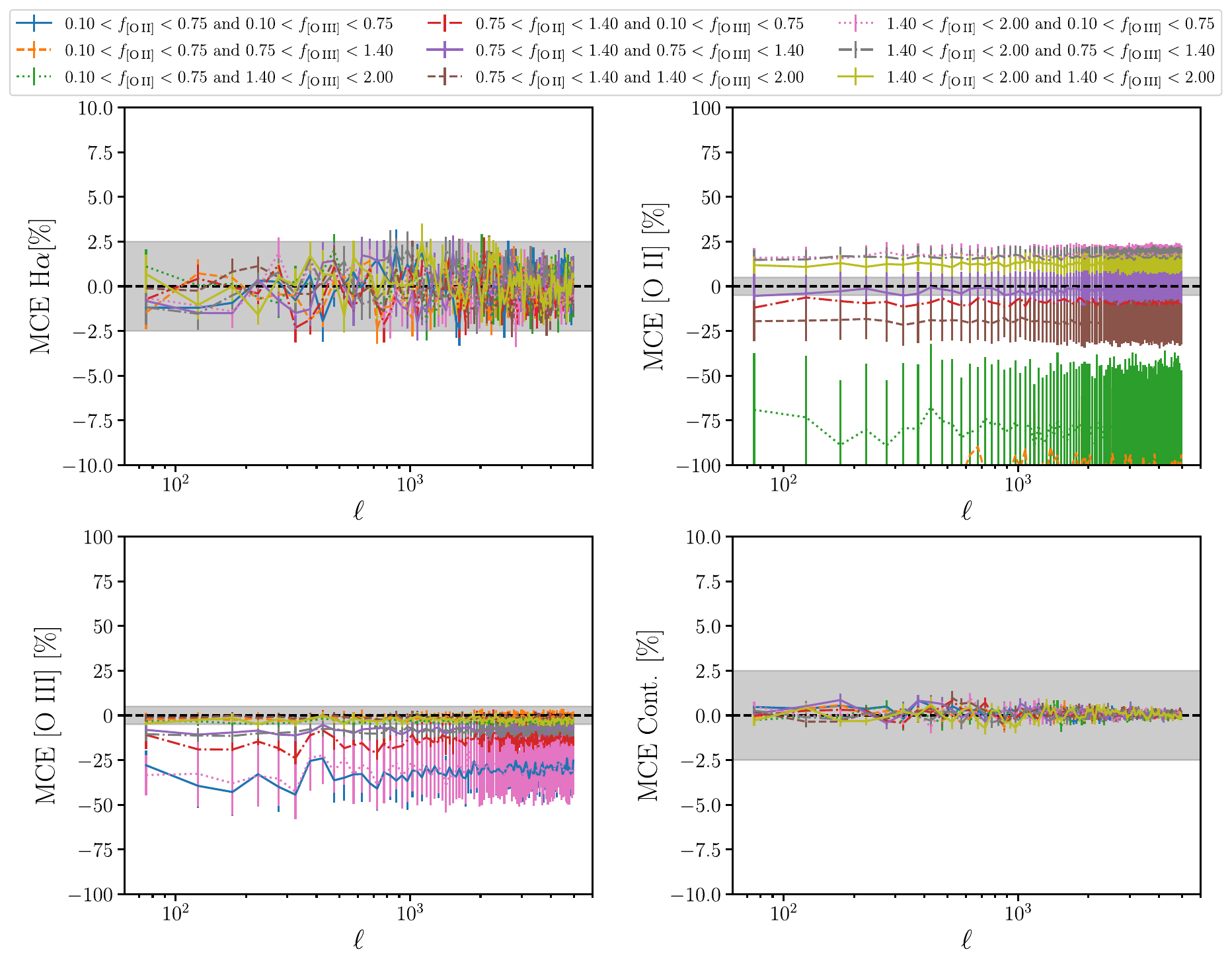}
    \caption{Same as figure~\ref{fig:all-S02-MCE-Mch} for the multi-channel analysis of the line-plus-continuum sample with reduced scatter of $0.1 \, {\rm dex}$.}\vspace{0.15in}
    \label{fig:all-S01-MCE-Mch}
\end{figure*}
\begin{figure*}
    \centering
    \includegraphics[width=0.9\linewidth]{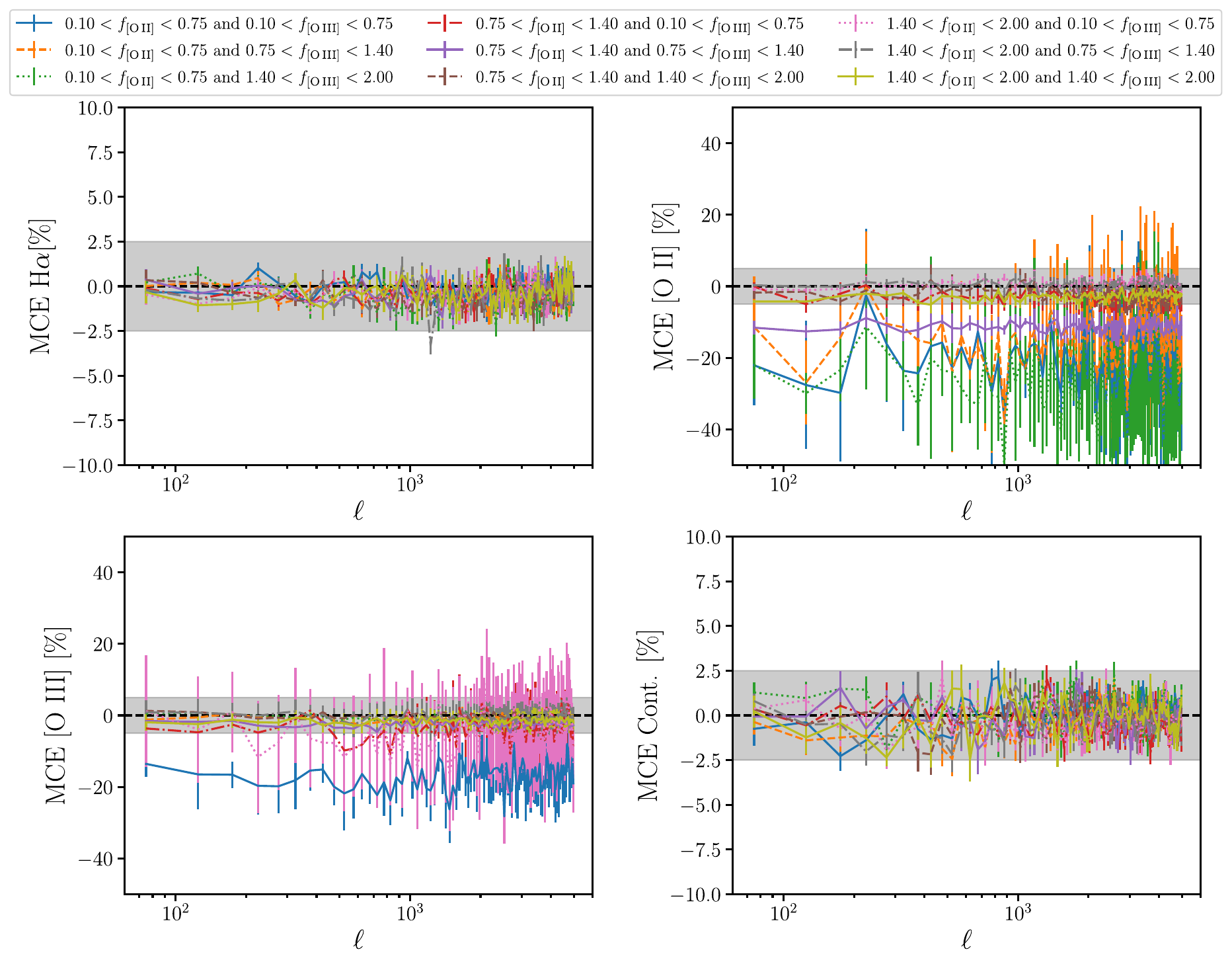}
    \caption{Same as figure~\ref{fig:all-S02-MCE-Mch} for the multi-channel analysis of the line-plus-continuum sample with reduced scatter and down scaled continuum.}\vspace{0.15in}
    \label{fig:all015-S01-MCE-Mch}
\end{figure*}
\begin{figure*}
    \centering
    \includegraphics[width=0.9\linewidth]{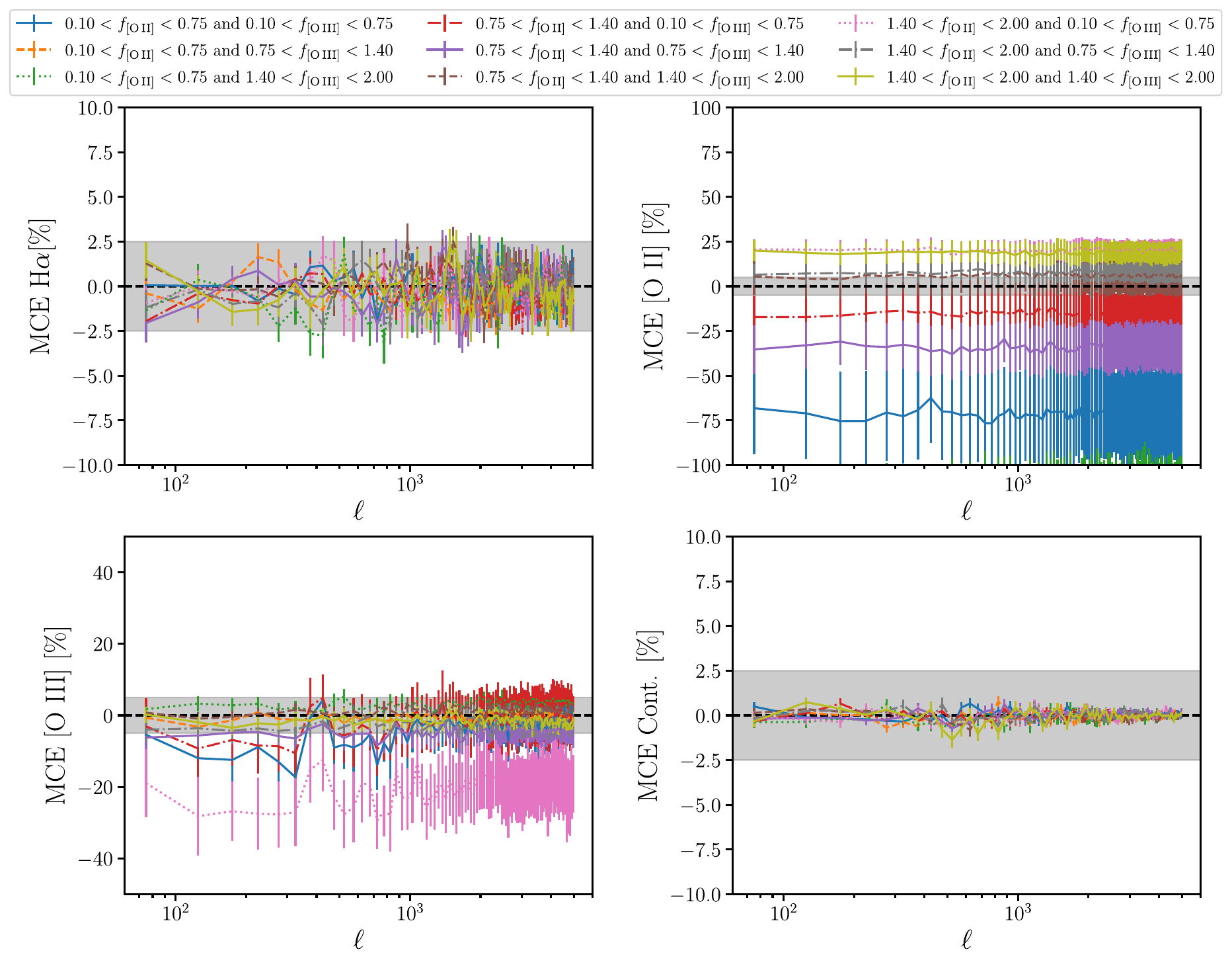}
    \caption{Same as figure~\ref{fig:all-S02-MCE-Mch} for the multi-channel plus cross continuum analysis of the line-plus-continuum sample with reduced scatter. }
    \label{fig:all-S01-MCE-MchxC}\vspace{0.15in}
\end{figure*}

\subsubsection{Scatter $(0.2 \, {\rm dex})$} 
\label{sec:scatter-lines}

The importance of the multi-channel information is even more evident when we add the log-normal scatter to the maps. In the second row of tables~\ref{tab:res-OII} and \ref{tab:res-OIII} we present the results for the line-only sample with $0.2 \, {\rm dex}$ scatter. At the level of the MSE, the multi-channel information improves the network's performance by two orders of magnitude for \oii and slightly more than one order for \oiii. As for the $\chi_{\rm red}^2$,  the multi-channel network is overconfident in its second-moment estimates (i.e., it underestimates the errors). By contrast, the single-channel $\chi_{\rm red}^2$ values are close to unity only because the inferred error bars are extremely large, not because the first-moment estimates are accurate, as the high MSE confirms.

The values of the scaling factor metrics for the single-channel network translate into its inability to recover the angular power spectrum for the different components; therefore, in figure~\ref{fig:line-S-MCE-Mch}, we present the mean correction error only for the multi-channel case. Although the network is overconfident in estimation of the error for the scaling factor, especially for \oii, this behavior is not evident in the angular power spectrum correction (see middle panel of figure~\ref{fig:line-S-MCE-Mch}). Adding the log-normal scatter substantially degrades the network's performance: the MSE increases by roughly one to two orders of magnitude, and residuals grow for all components. The recovery hierarchy mirrors the no-scatter case, but the typical residuals are now $< 2\%$ for \Ha, $< 5\%$ for \oiii (except in the lowest scaling-factor bin), and up to $\sim 30$ for \oii, which is most affected by the added scatter.

\subsection{Lines and continuum}
\label{sec:lines-plus-cont}

We next consider a more realistic setup in which the intensity maps include continuum emission. Given the results above, we restrict attention to the multi-channel input and do not analyze the single-channel case. We examine two scatter levels ($0.2$ and $0.1 \, {\rm dex}$), an augmented input that adds the cross–continuum spectrum, and a reduced-continuum scenario. Unless otherwise noted, all results below refer to the multi-channel network.

As described in section~\ref{sec:nointernet}, the network is trained to infer the contaminant scaling factors and to predict the corrections needed to recover, in the target channel, the angular power spectra of the target line, the interloper lines, and the continuum.

\subsubsection{Large scatter $(0.2 \, {\rm dex})$}
\label{sec:scatter02-cont}

We report MSE and for the line+continuum maps with scatter of $0.2 \, {\rm dex}$ in the third rows of Tables~\ref{tab:res-OII} and \ref{tab:res-OIII}. Relative to the corresponding lines-only analysis, the MSE degrades by roughly two orders of magnitude. In particular, the network fails to constrain the \oii scaling factor, reverting to the prior mean with appreciable dispersion. For \oiii, the mean scaling is recovered but with large scatter; however, the values are close to unity, indicating well–calibrated second-moment predictions in this configuration. This contrasts with the lines-only case, where uncertainty estimates were less reliable. 

In figure~\ref{fig:all-S02-MCE-Mch} we show the mean correction errors for all four components of the map. The hierarchical pattern persists: the brightest component (the continuum) is corrected most accurately, while \oii performs worst. The continuum and \Ha power spectra show no evident bias, with residuals $2.5\%$ and $6\%$, respectively. By contrast, for \oiii and \oii the network does not reliably recover the required correction, residuals remain substantially larger and can vary in sign across scaling–factor bins, indicating limited separability of the interloper lines in this configuration.

\subsubsection{Small scatter $(0.1 \, {\rm dex})$}
\label{sec:scatter01-cont}

For a lower scatter of $0.1 \, {\rm dex}$, we first present the results for the multi-channel network with two continuum levels in the maps (nominal and reduced), and then assess the impact of adding the cross-continuum input. The reduced continuum test rescales the continuum map by $0.15$, equivalent to a $0.02$ rescaling of the continuum angular power spectrum (section~\ref{sec:ch-choice}), and is intended to mimic PCA–cleaned maps with residual continuum.

As expected, the reduction in scatter greatly impacts the network's performance. The MSE of both \oii and \oiii improves by almost one order of magnitude. In particular, the network now infers the \oii scaling factor, unlike the previous case, even though there is still a very large scatter. At the level of the $\chi_{\rm red}^2$, the network tends to be more overconfident than before. The continuum rescaling brings another order of magnitude improvement in the performance, which is explained by the fact that the contaminant line emissions are no longer overwhelmed by the continuum. The network remains overconfident, especially in \oii.

As for the MCE, we observe a reduction of the residual error for all the components in figure~\ref{fig:all-S01-MCE-Mch} in comparison to figure~\ref{fig:all-S02-MCE-Mch}. The error is within $3\%$ and $1\%$ for \Ha and the continuum, respectively; while now we recover \oiii within $10\%$, when its scaling factor is not in the lowest bin. As for \oii, we can recover it within $25\%$ when, again, its scaling factor is not in its lower range. The more interesting comparison is with the results of the rescaled continuum case presented in figure~\ref{fig:all015-S01-MCE-Mch}. In this configuration, \Ha is the brightest component, and we can see that now its residual error is lower than the continuum one. Both components have a residual error within $2.5\%$, but \Ha has a smaller scatter. The reduced continuum level also positively affects the inference on the contaminant line corrections. The residual error is within $40\%$ for \oii (and $10\%$  if we remove the lower scaling factor bin), and $30\%$ for \oiii (and $5\%$ if we remove the lowest scaling factor bin).

The final configuration we tested is with the standard continuum level in the maps, but with an additional cross-correlation in the input. The new information we provide to the network should only pick the continuum angular power spectrum shape (see last paragraph of section~\ref{sec:ch-choice}). The MSE and $\chi_{\rm red}^2$ of this configuration are reported in the last column of Tables~\ref{tab:res-OII} and \ref{tab:res-OIII}. We observe a minimal improvement for \oii, but for \oiii the MSE is halved. As in the other configurations, the network is overconfident in its error estimation. 

We present the mean correction error for this final case in figure~\ref{fig:all-S01-MCE-MchxC}. Comparing with figure~\ref{fig:all-S01-MCE-Mch}, we see that the main improvement is for \oiii, which now has a residual error within $40\%$ or within $20\%$ if we exclude the combination of the high scaling factor range for \oii and the low one for \oiii. For the other components, the network correction ability is comparable with the case when we do not give the additional continuum cross-correlation as input. The network corrects the continuum within $1\%$, \Ha within $3\%$, and cannot properly correct \oii.

\section{Conclusions}
\label{sec:conclusions}

We investigated a neural–network approach, adapted from \citet{Cagliari:2025xcq}, to recover component angular power spectra from contaminated line–intensity maps. The network takes as input single– or multi–channel power–spectrum statistics and predicts the mean and spread of the scaling factors for interloper amplitudes and the corrections that map the contaminated spectrum in a target channel to the spectra of \Ha, \oii, \oiii, and (when present) the continuum.

We considered two data sets: first, maps with only the target and interloper lines, and second, maps that additionally include extragalactic continuum. In the line-only case, we first analyzed maps without added scatter, then added pixel-wise log-normal scatter to model astrophysical uncertainty and tested single- versus multi-channel inputs. We observed that multi-channel information is required to reach good correction performance. In the lines plus continuum case, we ran three analyses with multi-channel inputs: two scatter levels ($0.2$ and $0.1 \, {\rm dex}$) and a reduced-continuum scenario that mimics PCA cleaning (map rescale $0.15$). Across all configurations, we found a tendency to underestimate the uncertainties on the interloper scaling factors, and we interpret results conservatively.

In maps containing only \Ha, \oii, and \oiii, i.e., the line-only scenario, the residuals follow a brightness hierarchy: the brightest component is recovered most accurately. With single–channel input, the method achieves sub-percent recovery for \Ha that degrades for the interlopers. Adding multi–channel information improves performance by a factor of $\sim 4$ in the \Ha MCE and by $\sim 2-5$ for the interloper metrics; with pixel-wise log-normal scatter, the improvement grows to two orders of magnitude in MSE for \oii and slightly more than one order for \oiii. In the no–scatter case, the reduced chi-square, which is a diagnostic for second–moment estimates, shows configuration-dependent miscalibration, which largely disappears when scatter is included.

With continuum emission added, i.e., line plus continuum scenario, the brightness hierarchy persists. The continuum and \Ha spectra are recovered with typical residuals of $\leq 2.5\%$ and $\leq 6\%$, respectively, whereas the interloper spectra are not reliably recovered at nominal continuum levels. Reducing the continuum amplitude to mimic PCA cleaning yields interloper residuals of $\sim 10\%$ for \oii and  $\sim 5\%$ for \oiii (excluding the lowest scaling-factor bins), consistent with the reduced dynamic range between components. Adding cross-continuum information to the input further lowers the interloper-scaling MSE by $\sim 15\%$  for \oii and $\sim 40\%$ for \oiii, producing a visible improvement for \oiii.

Overall, our results demonstrate that the recovery accuracy is set by the relative brightness of the components, and multi–channel information is essential once realistic scatter and continuum are present. Even in the most challenging configuration tested (continuum plus interlopers with $0.2 \, {\rm dex}$ scatter), the \Ha spectrum in the target channel is recovered to $\sim 6\%$, improving to $\sim 3\%$ for reduced scatter of $0.1 \, {\rm dex}$.

This study represents a proof of concept. Before application to data, several extensions are needed: first, allow redshift-dependent interloper scalings and assess their impact; second, replace ad-hoc scaling factors with physically motivated nuisance parameters (e.g., metallicity or SFR prescriptions) and explore alternative uncertainty injections beyond pixel-wise log-normal scatter; and third, test robustness to cosmological variations and investigate whether multi-channel information helps break interloper–cosmology degeneracies, or whether additional statistics (e.g., three-point information) are required, as in the galaxy-clustering case.

\section*{Acknowledgments}
We thank Guochao Sun for the discussion related to physical modeling of scatter in LIM signal, and Yun-Ting Cheng and Martin White for their feedback on the first version of this manuscript. ZG thanks Olivier Dor{\'e}, Tzu-Ching Chang, Martin White and Chirag Modi for useful discussion in the construction of the simulation. This work is supported by the \emph{Agence Nationale de la Recherche} (ANR), grant 
ANR-23-CPJ1-0160-01. The LIM simulation for SPHEREx was produced on Perlmutter supercomputer of NERSC, and some of the post processing on the datasets for this project were done with computing and storage resources by GENCI at IDRIS, thanks to the grant 2025-AD010416295 on the CSL partition of the supercomputer Jean Zay. The analysis of this work has been done thanks to the facilities offered by the Univ. Savoie Mont Blanc - CNRS/IN2P3 MUST computing center. 

\bibliography{main}{}

\begin{thebibliography}{}
\expandafter\ifx\csname natexlab\endcsname\relax\def\natexlab#1{#1}\fi
\providecommand{\url}[1]{\href{#1}{#1}}
\providecommand{\dodoi}[1]{doi:~\href{http://doi.org/#1}{\nolinkurl{#1}}}
\providecommand{\doeprint}[1]{\href{http://ascl.net/#1}{\nolinkurl{http://ascl.net/#1}}}
\providecommand{\doarXiv}[1]{\href{https://arxiv.org/abs/#1}{\nolinkurl{https://arxiv.org/abs/#1}}}

\bibitem[{P.~A.~R. Ade {et~al.}(2020)Ade {et~al.}}]{Ade:2019ril}
Ade, P. A.~R., {et~al.} 2020, \bibinfo{title}{{The Experiment for Cryogenic Large-aperture Intensity Mapping (EXCLAIM)},} J. Low Temp. Phys., 199, 1027, \dodoi{10.1007/s10909-019-02320-5}

\bibitem[{P. Behroozi {et~al.}(2019)Behroozi, Wechsler, Hearin, \& Conroy}]{Behroozi_2019}
Behroozi, P., Wechsler, R.~H., Hearin, A.~P., \& Conroy, C. 2019, \bibinfo{title}{UniverseMachine: The correlation between galaxy growth and dark matter halo assembly from z = 0−10,} Monthly Notices of the Royal Astronomical Society, 488, 3143–3194, \dodoi{10.1093/mnras/stz1182}

\bibitem[{J.~L. Bernal {et~al.}(2019{\natexlab{a}})Bernal, Breysse, Gil-Mar{\'\i}n, \& Kovetz}]{Bernal:2019jdo}
Bernal, J.~L., Breysse, P.~C., Gil-Mar{\'\i}n, H., \& Kovetz, E.~D. 2019{\natexlab{a}}, \bibinfo{title}{{User{\textquoteright}s guide to extracting cosmological information from line-intensity maps},} Phys. Rev. D, 100, 123522, \dodoi{10.1103/PhysRevD.100.123522}

\bibitem[{J.~L. Bernal {et~al.}(2019{\natexlab{b}})Bernal, Breysse, \& Kovetz}]{Bernal:2019gfq}
Bernal, J.~L., Breysse, P.~C., \& Kovetz, E.~D. 2019{\natexlab{b}}, \bibinfo{title}{{Cosmic Expansion History from Line-Intensity Mapping},} Phys. Rev. Lett., 123, 251301, \dodoi{10.1103/PhysRevLett.123.251301}

\bibitem[{J.~L. Bernal {et~al.}(2021{\natexlab{a}})Bernal, Caputo, \& Kamionkowski}]{Bernal:2020lkd}
Bernal, J.~L., Caputo, A., \& Kamionkowski, M. 2021{\natexlab{a}}, \bibinfo{title}{{Strategies to Detect Dark-Matter Decays with Line-Intensity Mapping},} Phys. Rev. D, 103, 063523, \dodoi{10.1103/PhysRevD.103.063523}

\bibitem[{J.~L. Bernal {et~al.}(2021{\natexlab{b}})Bernal, Caputo, Villaescusa-Navarro, \& Kamionkowski}]{Bernal:2021ylz}
Bernal, J.~L., Caputo, A., Villaescusa-Navarro, F., \& Kamionkowski, M. 2021{\natexlab{b}}, \bibinfo{title}{{Searching for the Radiative Decay of the Cosmic Neutrino Background with Line-Intensity Mapping},} Phys. Rev. Lett., 127, 131102, \dodoi{10.1103/PhysRevLett.127.131102}

\bibitem[{J.~L. Bernal \& E.~D. Kovetz(2022)Bernal \& Kovetz}]{Bernal:2022jap}
Bernal, J.~L., \& Kovetz, E.~D. 2022, \bibinfo{title}{{Line-intensity mapping: theory review with a focus on star-formation lines},} Astron. Astrophys. Rev., 30, 5, \dodoi{10.1007/s00159-022-00143-0}

\bibitem[{M. Bethermin {et~al.}(2022)Bethermin {et~al.}}]{Bethermin:2022lmd}
Bethermin, M., {et~al.} 2022, \bibinfo{title}{{CONCERTO: High-fidelity simulation of millimeter line emissions of galaxies and [CII] intensity mapping},} Astron. Astrophys., 667, A156, \dodoi{10.1051/0004-6361/202243888}

\bibitem[{P.~C. {Breysse} \& R.~M. {Alexandroff}(2019){Breysse} \& {Alexandroff}}]{Breysse:2019MNRAS}
{Breysse}, P.~C., \& {Alexandroff}, R.~M. 2019, \bibinfo{title}{{Observing AGN feedback with CO intensity mapping},} \mnras, 490, 260, \dodoi{10.1093/mnras/stz2534}

\bibitem[{P.~C. Breysse {et~al.}(2016)Breysse, Kovetz, \& Kamionkowski}]{Breysse:2015saa}
Breysse, P.~C., Kovetz, E.~D., \& Kamionkowski, M. 2016, \bibinfo{title}{{The high redshift star-formation history from carbon-monoxide intensity maps},} Mon. Not. Roy. Astron. Soc., 457, L127, \dodoi{10.1093/mnrasl/slw005}

\bibitem[{P.~C. Breysse \& M. Rahman(2017)Breysse \& Rahman}]{Breysse:2016opl}
Breysse, P.~C., \& Rahman, M. 2017, \bibinfo{title}{{Feeding cosmic star formation: Exploring high-redshift molecular gas with CO intensity mapping},} Mon. Not. Roy. Astron. Soc., 468, 741, \dodoi{10.1093/mnras/stx451}

\bibitem[{P.~C. Breysse {et~al.}(2022)Breysse {et~al.}}]{COMAP:2021nrp}
Breysse, P.~C., {et~al.} 2022, \bibinfo{title}{{COMAP Early Science. VII. Prospects for CO Intensity Mapping at Reionization},} Astrophys. J., 933, 188, \dodoi{10.3847/1538-4357/ac63c9}

\bibitem[{T. {Brown} \& C.~D. {Wilson}(2019){Brown} \& {Wilson}}]{Brown:2019ApJ}
{Brown}, T., \& {Wilson}, C.~D. 2019, \bibinfo{title}{{Extreme CO Isotopologue Line Ratios in ULIRGS: Evidence for a Top-heavy IMF},} \apj, 879, 17, \dodoi{10.3847/1538-4357/ab2246}

\bibitem[{M.~S. Cagliari {et~al.}(2025)Cagliari, Moradinezhad~Dizgah, \& Villaescusa-Navarro}]{Cagliari:2025xcq}
Cagliari, M.~S., Moradinezhad~Dizgah, A., \& Villaescusa-Navarro, F. 2025, \bibinfo{title}{{Correcting for interloper contamination in the power spectrum with neural networks},} \doarXiv{2504.06919}

\bibitem[{T.-C. {Chang} {et~al.}(2010){Chang}, {Pen}, {Bandura}, \& {Peterson}}]{Chang2010}
{Chang}, T.-C., {Pen}, U.-L., {Bandura}, K., \& {Peterson}, J.~B. 2010, \bibinfo{title}{{An intensity map of hydrogen 21-cm emission at redshift z\raisebox{-0.5ex}\textasciitilde0.8},} Nature, 466, 463, \dodoi{10.1038/nature09187}

\bibitem[{T.-C. {Chang} {et~al.}(2019){Chang}, {Beane}, {Dore}, {Lidz}, {Mas-Ribas}, {Sun}, {Alvarez}, {Thakur}, {Berger}, {Bethermin}, {Bock}, {Bradford}, {Breysse}, {Burgarella}, {Charmandaris}, {Cheng}, {Cleary}, {Cooray}, {Crites}, {Ewall-Wice}, {Fan}, {Finkelstein}, {Furlanetto}, {Hewitt}, {Hunacek}, {Korngut}, {Kovetz}, {Hallinan}, {Heneka}, {Lagache}, {Lawrence}, {Lazio}, {Liu}, {Marrone}, {Parsons}, {Readhead}, {Rhodes}, {Riechers}, {Seiffert}, {Stacey}, {Visbal}, {Wu}, {Zemcov}, \& {Zheng}}]{Chang:2019BAAS}
{Chang}, T.-C., {Beane}, A., {Dore}, O., {et~al.} 2019, \bibinfo{title}{{Tomography of the Cosmic Dawn and Reionization Eras with Multiple Tracers},} \baas, 51, 282, \dodoi{10.48550/arXiv.1903.11744}

\bibitem[{E. {Chapman} {et~al.}(2012){Chapman}, {Abdalla}, {Harker}, {Jeli{\'c}}, {Labropoulos}, {Zaroubi}, {Brentjens}, {de Bruyn}, \& {Koopmans}}]{Chapman2012}
{Chapman}, E., {Abdalla}, F.~B., {Harker}, G., {et~al.} 2012, \bibinfo{title}{{Foreground removal using FASTICA: a showcase of LOFAR-EoR},} \mnras, 423, 2518, \dodoi{10.1111/j.1365-2966.2012.21065.x}

\bibitem[{S.~C. Chapman {et~al.}(2022)Chapman {et~al.}}]{CCAT-prime:2022qkj}
Chapman, S.~C., {et~al.} 2022, \bibinfo{title}{{CCAT-prime: the 850 GHz camera for prime-cam on FYST},} Proc. SPIE Int. Soc. Opt. Eng., 12190, 77, \dodoi{10.1117/12.2630628}

\bibitem[{C. Chen \& A.~R. Pullen(2022)Chen \& Pullen}]{Chen:2021ykb}
Chen, C., \& Pullen, A.~R. 2022, \bibinfo{title}{{Removing interlopers from intensity mapping probes of primordial non-Gaussianity},} Mon. Not. Roy. Astron. Soc., 512, 4262, \dodoi{10.1093/mnras/stac743}

\bibitem[{Y.-T. {Cheng} {et~al.}(2016){Cheng}, {Chang}, {Bock}, {Bradford}, \& {Cooray}}]{Cheng2016}
{Cheng}, Y.-T., {Chang}, T.-C., {Bock}, J., {Bradford}, C.~M., \& {Cooray}, A. 2016, \bibinfo{title}{{Spectral Line De-confusion in an Intensity Mapping Survey},} ApJ, 832, 165, \dodoi{10.3847/0004-637X/832/2/165}

\bibitem[{Y.-T. Cheng {et~al.}(2020)Cheng, Chang, \& Bock}]{Cheng:2020asz}
Cheng, Y.-T., Chang, T.-C., \& Bock, J.~J. 2020, \bibinfo{title}{{Phase-space Spectral Line Deconfusion in Intensity Mapping},} Astrophys. J., 901, 142, \dodoi{10.3847/1538-4357/abb023}

\bibitem[{Y.-T. Cheng {et~al.}(2023)Cheng, Wandelt, Chang, \& Dore}]{Cheng:2022ani}
Cheng, Y.-T., Wandelt, B.~D., Chang, T.-C., \& Dore, O. 2023, \bibinfo{title}{{Data-driven Cosmology from Three-dimensional Light Cones},} Astrophys. J., 944, 151, \dodoi{10.3847/1538-4357/acb350}

\bibitem[{Y.-T. Cheng {et~al.}(2024)Cheng, Wang, Wandelt, Chang, \& Dor{\'e}}]{Cheng:2024nfy}
Cheng, Y.-T., Wang, K., Wandelt, B.~D., Chang, T.-C., \& Dor{\'e}, O. 2024, \bibinfo{title}{{Bayesian Multi-line Intensity Mapping},} Astrophys. J., 971, 159, \dodoi{10.3847/1538-4357/ad57b9}

\bibitem[{K.~A. Cleary {et~al.}(2021)Cleary {et~al.}}]{Cleary:2021dsp}
Cleary, K.~A., {et~al.} 2021, \bibinfo{title}{{COMAP Early Science: I. Overview},} \dodoi{10.3847/1538-4357/ac63cc}

\bibitem[{P. Comaschi \& A. Ferrara(2016)Comaschi \& Ferrara}]{Comaschi:2015waa}
Comaschi, P., \& Ferrara, A. 2016, \bibinfo{title}{{Probing high-redshift galaxies with Ly$\alpha$ intensity mapping},} Mon. Not. Roy. Astron. Soc., 455, 725, \dodoi{10.1093/mnras/stv2339}

\bibitem[{P. Comaschi {et~al.}(2016)Comaschi, Yue, \& Ferrara}]{Comaschi:2016soe}
Comaschi, P., Yue, B., \& Ferrara, A. 2016, \bibinfo{title}{{Observational challenges in Ly{\ensuremath{\alpha}} intensity mapping},} Mon. Not. Roy. Astron. Soc., 463, 3193, \dodoi{10.1093/mnras/stw2198}

\bibitem[{ {CONCERTO Collaboration} {et~al.}(2020){CONCERTO Collaboration}, {Ade}, {Aravena}, {Barria}, {Beelen}, {Benoit}, {B{\'e}thermin}, {Bounmy}, {Bourrion}, {Bres}, {De Breuck}, {Calvo}, {Cao}, {Catalano}, {D{\'e}sert}, {Dur{\'a}n}, {Fasano}, {Fenouillet}, {Garcia}, {Garde}, {Goupy}, {Groppi}, {Hoarau}, {Lagache}, {Lambert}, {Leggeri}, {Levy-Bertrand}, {Mac{\'\i}as-P{\'e}rez}, {Mani}, {Marpaud}, {Mauskopf}, {Monfardini}, {Pisano}, {Ponthieu}, {Prieur}, {Roni}, {Roudier}, {Tourres}, \& {Tucker}}]{CONCERTO2020}
{CONCERTO Collaboration}, {Ade}, P., {Aravena}, M., {et~al.} 2020, \bibinfo{title}{{A wide field-of-view low-resolution spectrometer at APEX: Instrument design and scientific forecast},} \aap, 642, A60, \dodoi{10.1051/0004-6361/202038456}

\bibitem[{C. {Conroy} \& J.~E. {Gunn}(2010){Conroy} \& {Gunn}}]{Conroy_2010}
{Conroy}, C., \& {Gunn}, J.~E. 2010, \bibinfo{title}{{The Propagation of Uncertainties in Stellar Population Synthesis Modeling. III. Model Calibration, Comparison, and Evaluation},} ApJ, 712, 833, \dodoi{10.1088/0004-637X/712/2/833}

\bibitem[{C. Conroy {et~al.}(2009)Conroy, White, \& Gunn}]{Conroy_2009}
Conroy, C., White, M., \& Gunn, J.~E. 2009, \bibinfo{title}{THE PROPAGATION OF UNCERTAINTIES IN STELLAR POPULATION SYNTHESIS MODELING. II. THE CHALLENGE OF COMPARING GALAXY EVOLUTION MODELS TO OBSERVATIONS,} The Astrophysical Journal, 708, 58–70, \dodoi{10.1088/0004-637x/708/1/58}

\bibitem[{A. Cooray {et~al.}(2016)Cooray {et~al.}}]{Cooray:2016hro}
Cooray, A., {et~al.} 2016, \bibinfo{title}{{Cosmic Dawn Intensity Mapper},} \doarXiv{1602.05178}

\bibitem[{C. Creque-Sarbinowski \& M. Kamionkowski(2018)Creque-Sarbinowski \& Kamionkowski}]{Creque-Sarbinowski:2018ebl}
Creque-Sarbinowski, C., \& Kamionkowski, M. 2018, \bibinfo{title}{{Searching for Decaying and Annihilating Dark Matter with Line Intensity Mapping},} Phys. Rev. D, 98, 063524, \dodoi{10.1103/PhysRevD.98.063524}

\bibitem[{A.~T. {Crites} {et~al.}(2014){Crites}, {Bock}, {Bradford}, {Chang}, {Cooray}, {Duband}, {Gong}, {Hailey-Dunsheath}, {Hunacek}, {Koch}, {Li}, {O'Brient}, {Prouve}, {Shirokoff}, {Silva}, {Staniszewski}, {Uzgil}, \& {Zemcov}}]{TIME}
{Crites}, A.~T., {Bock}, J.~J., {Bradford}, C.~M., {et~al.} 2014, \bibinfo{title}{{The TIME-Pilot intensity mapping experiment},} in Society of Photo-Optical Instrumentation Engineers (SPIE) Conference Series, Vol. 9153, Millimeter, Submillimeter, and Far-Infrared Detectors and Instrumentation for Astronomy VII, ed. W.~S. {Holland} \& J.~{Zmuidzinas}, 91531W, \dodoi{10.1117/12.2057207}

\bibitem[{J. Delabrouille {et~al.}(2021)Delabrouille {et~al.}}]{Delabrouille:2019thj}
Delabrouille, J., {et~al.} 2021, \bibinfo{title}{{Microwave spectro-polarimetry of matter and radiation across space and time},} Exper. Astron., 51, 1471, \dodoi{10.1007/s10686-021-09721-z}

\bibitem[{T. {Di Matteo} {et~al.}(2002){Di Matteo}, {Perna}, {Abel}, \& {Rees}}]{Matteo2002}
{Di Matteo}, T., {Perna}, R., {Abel}, T., \& {Rees}, M.~J. 2002, \bibinfo{title}{{Radio Foregrounds for the 21 Centimeter Tomography of the Neutral Intergalactic Medium at High Redshifts},} \apj, 564, 576, \dodoi{10.1086/324293}

\bibitem[{O. Doré {et~al.}(2018)Doré, Werner, Ashby, Bleem, Bock, Burt, Capak, Chang, Chaves-Montero, Chen, Civano, Cleeves, Cooray, Crill, Crossfield, Cushing, de~la Torre, DiMatteo, Dvory, Dvorkin, Espaillat, Ferraro, Finkbeiner, Greene, Hewitt, Hogg, Huffenberger, Jun, Ilbert, Jeong, Johnson, Kim, Kirkpatrick, Kowalski, Korngut, Li, Lisse, MacGregor, Mamajek, Mauskopf, Melnick, Ménard, Neyrinck, Öberg, Pisani, Rocca, Salvato, Schaan, Scoville, Song, Stevens, Tenneti, Teplitz, Tolls, Unwin, Urry, Wandelt, Williams, Wilner, Windhorst, Wolk, Yorke, \& Zemcov}]{Dore2018}
Doré, O., Werner, M.~W., Ashby, M. L.~N., {et~al.} 2018, Science Impacts of the SPHEREx All-Sky Optical to Near-Infrared Spectral Survey II: Report of a Community Workshop on the Scientific Synergies Between the SPHEREx Survey and Other Astronomy Observatories, arXiv, \dodoi{10.48550/ARXIV.1805.05489}

\bibitem[{R.~M. Feder {et~al.}(2025)Feder {et~al.}}]{CIBER:2025aoi}
Feder, R.~M., {et~al.} 2025, \bibinfo{title}{{CIBER 4th flight fluctuation analysis: Measurements of near-IR auto- and cross-power spectra on arcminute to sub-degree scales},} \doarXiv{2501.17933}

\bibitem[{Z. Gao {et~al.}(in prep.)Gao, Dor{\'e}, \& Chang}]{Gao-prep}
Gao, Z., Dor{\'e}, O., \& Chang, T. in prep., \bibinfo{title}{{A realistic study of continuum contamination to line intensity mapping with SPHEREx},} {A realistic study of continuum contamination to line intensity mapping with SPHEREx}

\bibitem[{K. Gebhardt {et~al.}(2021)Gebhardt {et~al.}}]{Gebhardt:2021vfo}
Gebhardt, K., {et~al.} 2021, \bibinfo{title}{{The Hobby{\textendash}Eberly Telescope Dark Energy Experiment (HETDEX) Survey Design, Reductions, and Detections*},} Astrophys. J., 923, 217, \dodoi{10.3847/1538-4357/ac2e03}

\bibitem[{Y. {Gong} {et~al.}(2012){Gong}, {Cooray}, {Silva}, {Santos}, {Bock}, {Bradford}, \& {Zemcov}}]{Gong2012}
{Gong}, Y., {Cooray}, A., {Silva}, M., {et~al.} 2012, \bibinfo{title}{{Intensity Mapping of the [C II] Fine Structure Line during the Epoch of Reionization},} \apj, 745, 49, \dodoi{10.1088/0004-637X/745/1/49}

\bibitem[{Y. Gong {et~al.}(2017)Gong, Cooray, Silva, Zemcov, Feng, Santos, Dore, \& Chen}]{Gong_2017}
Gong, Y., Cooray, A., Silva, M.~B., {et~al.} 2017, \bibinfo{title}{Intensity Mapping of H$\alpha$, H$\beta$, [O II], and [O III] Lines at z < 5,} The Astrophysical Journal, 835, 273, \dodoi{10.3847/1538-4357/835/2/273}

\bibitem[{N. Jeffrey \& B.~D. Wandelt(2020)Jeffrey \& Wandelt}]{Jeffrey:2020itg}
Jeffrey, N., \& Wandelt, B.~D. 2020, \bibinfo{title}{{Solving high-dimensional parameter inference: marginal posterior densities \& Moment Networks},} in {34th Conference on Neural Information Processing Systems}.
\newblock \doarXiv{2011.05991}

\bibitem[{K.~S. Karkare \& S. Bird(2018)Karkare \& Bird}]{Karkare:2018sar}
Karkare, K.~S., \& Bird, S. 2018, \bibinfo{title}{{Constraining the Expansion History and Early Dark Energy with Line Intensity Mapping},} Phys. Rev. D, 98, 043529, \dodoi{10.1103/PhysRevD.98.043529}

\bibitem[{K.~S. Karkare {et~al.}(2022{\natexlab{a}})Karkare, Moradinezhad~Dizgah, Keating, Breysse, \& Chung}]{Karkare:2022bai}
Karkare, K.~S., Moradinezhad~Dizgah, A., Keating, G.~K., Breysse, P., \& Chung, D.~T. 2022{\natexlab{a}}, \bibinfo{title}{{Snowmass 2021 Cosmic Frontier White Paper: Cosmology with Millimeter-Wave Line Intensity Mapping},} in {Snowmass 2021}.
\newblock \doarXiv{2203.07258}

\bibitem[{K.~S. Karkare {et~al.}(2022{\natexlab{b}})Karkare {et~al.}}]{Karkare:2021ryi}
Karkare, K.~S., {et~al.} 2022{\natexlab{b}}, \bibinfo{title}{{SPT-SLIM: A Line Intensity Mapping Pathfinder for the South Pole Telescope},} J. Low Temp. Phys., 209, 758, \dodoi{10.1007/s10909-022-02702-2}

\bibitem[{G.~K. Keating {et~al.}(2020)Keating, Marrone, Bower, \& Keenan}]{Keating:2020wlx}
Keating, G.~K., Marrone, D.~P., Bower, G.~C., \& Keenan, R.~P. 2020, \bibinfo{title}{{An Intensity Mapping Detection of Aggregate CO Line Emission at 3 mm},} Astrophys. J., 901, 141, \dodoi{10.3847/1538-4357/abb08e}

\bibitem[{G.~K. Keating {et~al.}(2016)Keating, Marrone, Bower, Leitch, Carlstrom, \& DeBoer}]{Keating:2016pka}
Keating, G.~K., Marrone, D.~P., Bower, G.~C., {et~al.} 2016, \bibinfo{title}{{COPSS II: The molecular gas content of ten million cubic megaparsecs at redshift z = 3},} Astrophys. J., 830, 34, \dodoi{10.3847/0004-637X/830/1/34}

\bibitem[{D.~P. Kingma \& J. Ba(2014)Kingma \& Ba}]{kingma2014adam}
Kingma, D.~P., \& Ba, J. 2014, \bibinfo{title}{Adam: A method for stochastic optimization,} \doarXiv{1412.6980}

\bibitem[{E. {Kovetz} {et~al.}(2019){Kovetz}, {Breysse}, {Lidz}, {Bock}, {Bradford}, {Chang}, {Foreman}, {Padmanabhan}, {Pullen}, {Riechers}, {Silva}, \& {Switzer}}]{Kovetz:2019BAAS}
{Kovetz}, E., {Breysse}, P.~C., {Lidz}, A., {et~al.} 2019, \bibinfo{title}{{Astrophysics and Cosmology with Line-Intensity Mapping},} \baas, 51, 101, \dodoi{10.48550/arXiv.1903.04496}

\bibitem[{E.~D. Kovetz {et~al.}(2017)Kovetz {et~al.}}]{Kovetz:2017agg}
Kovetz, E.~D., {et~al.} 2017, \bibinfo{title}{{Line-Intensity Mapping: 2017 Status Report},} \doarXiv{1709.09066}

\bibitem[{S. Lee {et~al.}(in prep.)Lee {et~al.}}]{Lee-prep}
Lee, S., {et~al.} in prep., \bibinfo{title}{{Euclid preparation. TBD. The impact of redshift interlopers on the power spectrum},} {Euclid preparation. TBD. The impact of redshift interlopers on the power spectrum}

\bibitem[{T.~Y. Li {et~al.}(2016)Li, Wechsler, Devaraj, \& Church}]{Li:2015gqa}
Li, T.~Y., Wechsler, R.~H., Devaraj, K., \& Church, S.~E. 2016, \bibinfo{title}{{Connecting CO Intensity Mapping to Molecular Gas and Star Formation in the Epoch of Galaxy Assembly},} Astrophys. J., 817, 169, \dodoi{10.3847/0004-637X/817/2/169}

\bibitem[{A. {Lidz} \& J. {Taylor}(2016){Lidz} \& {Taylor}}]{Lidz2016}
{Lidz}, A., \& {Taylor}, J. 2016, \bibinfo{title}{{On Removing Interloper Contamination from Intensity Mapping Power Spectrum Measurements},} \apj, 825, 143, \dodoi{10.3847/0004-637X/825/2/143}

\bibitem[{A. Liu \& J.~R. Shaw(2020)Liu \& Shaw}]{Liu:2019awk}
Liu, A., \& Shaw, J.~R. 2020, \bibinfo{title}{{Data Analysis for Precision 21 cm Cosmology},} Publ. Astron. Soc. Pac., 132, 062001, \dodoi{10.1088/1538-3873/ab5bfd}

\bibitem[{A. {Liu} \& M. {Tegmark}(2011){Liu} \& {Tegmark}}]{Liu2011}
{Liu}, A., \& {Tegmark}, M. 2011, \bibinfo{title}{{A method for 21 cm power spectrum estimation in the presence of foregrounds},} \prd, 83, 103006, \dodoi{10.1103/PhysRevD.83.103006}

\bibitem[{A. {Liu} {et~al.}(2009){Liu}, {Tegmark}, {Bowman}, {Hewitt}, \& {Zaldarriaga}}]{Liu2009}
{Liu}, A., {Tegmark}, M., {Bowman}, J., {Hewitt}, J., \& {Zaldarriaga}, M. 2009, \bibinfo{title}{{An improved method for 21-cm foreground removal},} \mnras, 398, 401, \dodoi{10.1111/j.1365-2966.2009.15156.x}

\bibitem[{Y. {Liu} {et~al.}(2013){Liu}, {Madlener}, {Wolf}, \& {Wang}}]{Liu:2013}
{Liu}, Y., {Madlener}, D., {Wolf}, S., \& {Wang}, H.-C. 2013, \bibinfo{title}{{A comparison of approaches in fitting continuum SEDs},} Research in Astronomy and Astrophysics, 13, 420, \dodoi{10.1088/1674-4527/13/4/005}

\bibitem[{A.~L. Maas(2013)Maas}]{Maas2013RectifierNI}
Maas, A.~L. 2013, \bibinfo{title}{Rectifier Nonlinearities Improve Neural Network Acoustic Models,} in Semantic Scholar.
\newblock \url{https://api.semanticscholar.org/CorpusID:16489696}

\bibitem[{N. {Mashian} {et~al.}(2015){Mashian}, {Sternberg}, \& {Loeb}}]{Mashian:2015JCAP}
{Mashian}, N., {Sternberg}, A., \& {Loeb}, A. 2015, \bibinfo{title}{{Predicting the intensity mapping signal for multi-J CO lines},} \jcap, 2015, 028, \dodoi{10.1088/1475-7516/2015/11/028}

\bibitem[{M. {McQuinn} {et~al.}(2006){McQuinn}, {Zahn}, {Zaldarriaga}, {Hernquist}, \& {Furlanetto}}]{McQuinn2006}
{McQuinn}, M., {Zahn}, O., {Zaldarriaga}, M., {Hernquist}, L., \& {Furlanetto}, S.~R. 2006, \bibinfo{title}{{Cosmological Parameter Estimation Using 21 cm Radiation from the Epoch of Reionization},} \apj, 653, 815, \dodoi{10.1086/505167}

\bibitem[{C. Modi {et~al.}(2019)Modi, Castorina, Feng, \& White}]{Modi2019}
Modi, C., Castorina, E., Feng, Y., \& White, M. 2019, \bibinfo{title}{Intensity mapping with neutral hydrogen and the Hidden Valley simulations,} Journal of Cosmology and Astroparticle Physics, 2019, 024–024, \dodoi{10.1088/1475-7516/2019/09/024}

\bibitem[{A. Moradinezhad~Dizgah {et~al.}(2024)Moradinezhad~Dizgah, Bellini, \& Keating}]{MoradinezhadDizgah:2023src}
Moradinezhad~Dizgah, A., Bellini, E., \& Keating, G.~K. 2024, \bibinfo{title}{{Probing Dark Energy and Modifications of Gravity with Ground-based millimeter-wavelength Line Intensity Mapping},} Astrophys. J., 965, 19, \dodoi{10.3847/1538-4357/ad2078}

\bibitem[{A. Moradinezhad~Dizgah \& G.~K. Keating(2019)Moradinezhad~Dizgah \& Keating}]{MoradinezhadDizgah:2018lac}
Moradinezhad~Dizgah, A., \& Keating, G.~K. 2019, \bibinfo{title}{{Line intensity mapping with [CII] and CO(1-0) as probes of primordial non-Gaussianity},} Astrophys. J., 872, 126, \dodoi{10.3847/1538-4357/aafd36}

\bibitem[{A. Moradinezhad~Dizgah {et~al.}(2019)Moradinezhad~Dizgah, Keating, \& Fialkov}]{MoradinezhadDizgah:2018zrs}
Moradinezhad~Dizgah, A., Keating, G.~K., \& Fialkov, A. 2019, \bibinfo{title}{{Probing Cosmic Origins with CO and [CII] Emission Lines},} Astrophys. J. Lett., 870, L4, \dodoi{10.3847/2041-8213/aaf813}

\bibitem[{A. Moradinezhad~Dizgah {et~al.}(2022{\natexlab{a}})Moradinezhad~Dizgah, Keating, Karkare, Crites, \& Choudhury}]{MoradinezhadDizgah:2021upg}
Moradinezhad~Dizgah, A., Keating, G.~K., Karkare, K.~S., Crites, A., \& Choudhury, S.~R. 2022{\natexlab{a}}, \bibinfo{title}{{Neutrino Properties with Ground-based Millimeter-wavelength Line Intensity Mapping},} Astrophys. J., 926, 137, \dodoi{10.3847/1538-4357/ac3edd}

\bibitem[{A. Moradinezhad~Dizgah {et~al.}(2022{\natexlab{b}})Moradinezhad~Dizgah, Nikakhtar, Keating, \& Castorina}]{MoradinezhadDizgah:2021dei}
Moradinezhad~Dizgah, A., Nikakhtar, F., Keating, G.~K., \& Castorina, E. 2022{\natexlab{b}}, \bibinfo{title}{{Precision tests of CO and CII power spectra models against simulated intensity maps},} JCAP, 02, 026, \dodoi{10.1088/1475-7516/2022/02/026}

\bibitem[{S.~P. {Oh} \& K.~J. {Mack}(2003){Oh} \& {Mack}}]{Oh2003}
{Oh}, S.~P., \& {Mack}, K.~J. 2003, \bibinfo{title}{{Foregrounds for 21-cm observations of neutral gas at high redshift},} \mnras, 346, 871, \dodoi{10.1111/j.1365-2966.2003.07133.x}

\bibitem[{J.~R. Pritchard \& A. Loeb(2012)Pritchard \& Loeb}]{Pritchard:2011xb}
Pritchard, J.~R., \& Loeb, A. 2012, \bibinfo{title}{{21-cm cosmology},} Rept. Prog. Phys., 75, 086901, \dodoi{10.1088/0034-4885/75/8/086901}

\bibitem[{A.~R. {Pullen} {et~al.}(2013){Pullen}, {Chang}, {Dor{\'e}}, \& {Lidz}}]{Pullen2013}
{Pullen}, A.~R., {Chang}, T.-C., {Dor{\'e}}, O., \& {Lidz}, A. 2013, \bibinfo{title}{{Cross-correlations as a Cosmological Carbon Monoxide Detector},} \apj, 768, 15, \dodoi{10.1088/0004-637X/768/1/15}

\bibitem[{A.~R. Pullen {et~al.}(2014)Pullen, Dore, \& Bock}]{Pullen:2013dir}
Pullen, A.~R., Dore, O., \& Bock, J. 2014, \bibinfo{title}{{Intensity Mapping across Cosmic Times with the $Ly\alpha$ Line},} Astrophys. J., 786, 111, \dodoi{10.1088/0004-637X/786/2/111}

\bibitem[{M. {Raissi} {et~al.}(2019){Raissi}, {Perdikaris}, \& {Karniadakis}}]{2019JCoPh.378..686R}
{Raissi}, M., {Perdikaris}, P., \& {Karniadakis}, G.~E. 2019, \bibinfo{title}{{Physics-informed neural networks: A deep learning framework for solving forward and inverse problems involving nonlinear partial differential equations},} Journal of Computational Physics, 378, 686, \dodoi{10.1016/j.jcp.2018.10.045}

\bibitem[{I. Risso {et~al.}(2025)Risso {et~al.}}]{Euclid:2025duk}
Risso, I., {et~al.} 2025, \bibinfo{title}{{Euclid preparation. The impact of redshift interlopers on the two-point correlation function analysis},} \doarXiv{2505.04688}

\bibitem[{A. Roy \& N. Battaglia(2024)Roy \& Battaglia}]{Roy:2023pei}
Roy, A., \& Battaglia, N. 2024, \bibinfo{title}{{Cross-correlation Techniques to Mitigate the Interloper Contamination for Line Intensity Mapping Experiments},} Astrophys. J., 969, 2, \dodoi{10.3847/1538-4357/ad4a29}

\bibitem[{M.~G. {Santos} {et~al.}(2005){Santos}, {Cooray}, \& {Knox}}]{Santos2005}
{Santos}, M.~G., {Cooray}, A., \& {Knox}, L. 2005, \bibinfo{title}{{Multifrequency Analysis of 21 Centimeter Fluctuations from the Era of Reionization},} \apj, 625, 575, \dodoi{10.1086/429857}

\bibitem[{G. Sato-Polito {et~al.}(2020)Sato-Polito, Bernal, Kovetz, \& Kamionkowski}]{Sato-Polito:2020qpc}
Sato-Polito, G., Bernal, J.~L., Kovetz, E.~D., \& Kamionkowski, M. 2020, \bibinfo{title}{{Antisymmetric cross-correlation of line-intensity maps as a probe of reionization},} Phys. Rev. D, 102, 043519, \dodoi{10.1103/PhysRevD.102.043519}

\bibitem[{E. Schaan \& M. White(2021{\natexlab{a}})Schaan \& White}]{Schaan:2021gzb}
Schaan, E., \& White, M. 2021{\natexlab{a}}, \bibinfo{title}{{Multi-tracer intensity mapping: Cross-correlations, Line noise {\&} Decorrelation},} JCAP, 05, 068, \dodoi{10.1088/1475-7516/2021/05/068}

\bibitem[{E. Schaan \& M. White(2021{\natexlab{b}})Schaan \& White}]{Schaan:2021hhy}
Schaan, E., \& White, M. 2021{\natexlab{b}}, \bibinfo{title}{{Astrophysics {\&} Cosmology from Line Intensity Mapping vs Galaxy Surveys},} JCAP, 05, 067, \dodoi{10.1088/1475-7516/2021/05/067}

\bibitem[{B.~R. Scott {et~al.}(2023)Scott, Karkare, \& Bird}]{Scott:2022fev}
Scott, B.~R., Karkare, K.~S., \& Bird, S. 2023, \bibinfo{title}{{A forecast for large-scale structure constraints on Horndeski gravity with CO line intensity mapping},} Mon. Not. Roy. Astron. Soc., 523, 4895, \dodoi{10.1093/mnras/stad1501}

\bibitem[{G. Shmueli {et~al.}(2025)Shmueli, Libanore, \& Kovetz}]{Shmueli:2024npx}
Shmueli, G., Libanore, S., \& Kovetz, E.~D. 2025, \bibinfo{title}{{Toward a multitracer neutrino mass measurement with line-intensity mapping},} Phys. Rev. D, 111, 063512, \dodoi{10.1103/PhysRevD.111.063512}

\bibitem[{M. {Silva} {et~al.}(2015){Silva}, {Santos}, {Cooray}, \& {Gong}}]{Silva2015}
{Silva}, M., {Santos}, M.~G., {Cooray}, A., \& {Gong}, Y. 2015, \bibinfo{title}{{Prospects for Detecting C II Emission during the Epoch of Reionization},} \apj, 806, 209, \dodoi{10.1088/0004-637X/806/2/209}

\bibitem[{M.~B. Silva {et~al.}(2015)Silva, Santos, Cooray, \& Gong}]{Silva:2014ira}
Silva, M.~B., Santos, M.~G., Cooray, A., \& Gong, Y. 2015, \bibinfo{title}{{Prospects for Detecting C$\scriptsize{II}$ Emission During the Epoch of Reionization},} Astrophys. J., 806, 209, \dodoi{10.1088/0004-637X/806/2/209}

\bibitem[{M.~B. {Silva} {et~al.}(2013{\natexlab{a}}){Silva}, {Santos}, {Gong}, {Cooray}, \& {Bock}}]{Silva:2013ApJ}
{Silva}, M.~B., {Santos}, M.~G., {Gong}, Y., {Cooray}, A., \& {Bock}, J. 2013{\natexlab{a}}, \bibinfo{title}{{Intensity Mapping of Ly{\ensuremath{\alpha}} Emission during the Epoch of Reionization},} \apj, 763, 132, \dodoi{10.1088/0004-637X/763/2/132}

\bibitem[{M.~B. {Silva} {et~al.}(2013{\natexlab{b}}){Silva}, {Santos}, {Gong}, {Cooray}, \& {Bock}}]{Silva2013}
{Silva}, M.~B., {Santos}, M.~G., {Gong}, Y., {Cooray}, A., \& {Bock}, J. 2013{\natexlab{b}}, \bibinfo{title}{{Intensity Mapping of Ly{\ensuremath{\alpha}} Emission during the Epoch of Reionization},} \apj, 763, 132, \dodoi{10.1088/0004-637X/763/2/132}

\bibitem[{M.~B. Silva {et~al.}(2021)Silva, Kovetz, Keating, Moradinezhad~Dizgah, Bethermin, Breysse, Karkare, Bernal, \& Delabrouille}]{Silva:2019hsh}
Silva, M.~B., Kovetz, E.~D., Keating, G.~K., {et~al.} 2021, \bibinfo{title}{{Mapping large-scale-structure evolution over cosmic times},} Exper. Astron., 51, 1593, \dodoi{10.1007/s10686-021-09755-3}

\bibitem[{G. {Sun}(2022){Sun}}]{Sun2022}
{Sun}, G. 2022, \bibinfo{title}{{Cosmological Constraints on the Global Star Formation Law of Galaxies: Insights From Baryon Acoustic Oscillation Intensity Mapping},} arXiv e-prints, arXiv:2205.09354.
\newblock \doarXiv{2205.09354}

\bibitem[{G. {Sun} {et~al.}(2018){Sun}, {Moncelsi}, {Viero}, {Silva}, {Bock}, {Bradford}, {Chang}, {Cheng}, {Cooray}, {Crites}, {Hailey-Dunsheath}, {Uzgil}, {Hunacek}, \& {Zemcov}}]{Sun2018}
{Sun}, G., {Moncelsi}, L., {Viero}, M.~P., {et~al.} 2018, \bibinfo{title}{{A Foreground Masking Strategy for [C II] Intensity Mapping Experiments Using Galaxies Selected by Stellar Mass and Redshift},} \apj, 856, 107, \dodoi{10.3847/1538-4357/aab3e3}

\bibitem[{M. Van~Cuyck {et~al.}(2023)Van~Cuyck {et~al.}}]{VanCuyck:2023uli}
Van~Cuyck, M., {et~al.} 2023, \bibinfo{title}{{CONCERTO: Extracting the power spectrum of the [C II ] emission line},} Astron. Astrophys., 676, A62, \dodoi{10.1051/0004-6361/202346270}

\bibitem[{J. Vieira {et~al.}(2020)Vieira, Aguirre, Bradford, Filippini, Groppi, Marrone, Bethermin, Chang, Devlin, Dore, Fu, Dunsheath, Holder, Keating, Keenan, Kovetz, Lagache, Mauskopf, Narayanan, Popping, Shirokoff, Somerville, Trumper, Uzgil, \& Zmuidzinas}]{TIM}
Vieira, J., Aguirre, J., Bradford, C.~M., {et~al.} 2020, The Terahertz Intensity Mapper (TIM): a Next-Generation Experiment for Galaxy Evolution Studies, \doarXiv{2009.14340}

\bibitem[{F. Villaescusa-Navarro {et~al.}(2022)Villaescusa-Navarro {et~al.}}]{CAMELS:2021raw}
Villaescusa-Navarro, F., {et~al.} 2022, \bibinfo{title}{{The CAMELS Multifield Data Set: Learning the Universe\textquoteright{}s Fundamental Parameters with Artificial Intelligence},} Astrophys. J. Supp., 259, 61, \dodoi{10.3847/1538-4365/ac5ab0}

\bibitem[{E. {Visbal} \& A. {Loeb}(2010){Visbal} \& {Loeb}}]{Visbal2010}
{Visbal}, E., \& {Loeb}, A. 2010, \bibinfo{title}{{Measuring the 3D clustering of undetected galaxies through cross correlation of their cumulative flux fluctuations from multiple spectral lines},} JCAP, 2010, 016, \dodoi{10.1088/1475-7516/2010/11/016}

\bibitem[{S. Yang {et~al.}(2019)Yang, Pullen, \& Switzer}]{Yang:2019eoj}
Yang, S., Pullen, A.~R., \& Switzer, E.~R. 2019, \bibinfo{title}{{Evidence for C II diffuse line emission at redshift $z\sim2.6$},} Mon. Not. Roy. Astron. Soc., 489, L53, \dodoi{10.1093/mnrasl/slz126}

\bibitem[{B. Yue {et~al.}(2015)Yue, Ferrara, Pallottini, Gallerani, \& Vallini}]{Yue:2015sua}
Yue, B., Ferrara, A., Pallottini, A., Gallerani, S., \& Vallini, L. 2015, \bibinfo{title}{{Intensity mapping of [C II] emission from early galaxies},} Mon. Not. Roy. Astron. Soc., 450, 3829, \dodoi{10.1093/mnras/stv933}

\bibitem[{M. {Zaldarriaga} {et~al.}(2004){Zaldarriaga}, {Furlanetto}, \& {Hernquist}}]{Zaldarriaga2004}
{Zaldarriaga}, M., {Furlanetto}, S.~R., \& {Hernquist}, L. 2004, \bibinfo{title}{{21 Centimeter Fluctuations from Cosmic Gas at High Redshifts},} \apj, 608, 622, \dodoi{10.1086/386327}

\bibitem[{M. {Zhou} {et~al.}(2020){Zhou}, {Tan}, \& {Mao}}]{Zhou:2020arXiv}
{Zhou}, M., {Tan}, J., \& {Mao}, Y. 2020, \bibinfo{title}{{Robust Intensity Mapping Analysis against Foregrounds for the Epoch of Reionization},} arXiv e-prints, arXiv:2009.02765, \dodoi{10.48550/arXiv.2009.02765}

\bibitem[{M. Zhou {et~al.}(2021)Zhou, Tan, \& Mao}]{Zhou:2020hqh}
Zhou, M., Tan, J., \& Mao, Y. 2021, \bibinfo{title}{{Antisymmetric Cross-correlation between H I and CO Line Intensity Maps as a New Probe of Cosmic Reionization},} Astrophys. J., 909, 51, \dodoi{10.3847/1538-4357/abda45}

\end{thebibliography}
\bibliographystyle{aasjournalv7}


\end{document}